\documentclass[a4paper,12pt]{report}

\usepackage{amsmath} 
\usepackage{amssymb} 
\usepackage{graphicx} 
\usepackage{subcaption}
\usepackage{amsthm}

\usepackage{verbatim}
\usepackage{mathtools}
\usepackage{multirow}
\usepackage{todonotes}
\usepackage{booktabs}
\newcommand{\mycomment}[1]{}
\usepackage[UKenglish]{babel}

\usepackage[backref=section]{hyperref}

\hypersetup{
    citecolor=black,
    filecolor=black,
    linkcolor=green,
    urlcolor=black
}
\usepackage{flexisym}
\usepackage{listings}
\usepackage{color}
\usepackage{breakcites}
\usepackage{float}
\usepackage{mfirstuc}
\usepackage[showframe=false]{geometry}
\usepackage[utf8]{inputenc}
\definecolor{dkgreen}{rgb}{0,0.6,0}
\definecolor{gray}{rgb}{0.5,0.5,0.5}
\definecolor{mauve}{rgb}{0.58,0,0.82}

\begin{document}

\begin{titlepage}

\center 


\vspace{-1cm}
\includegraphics[width=4cm]{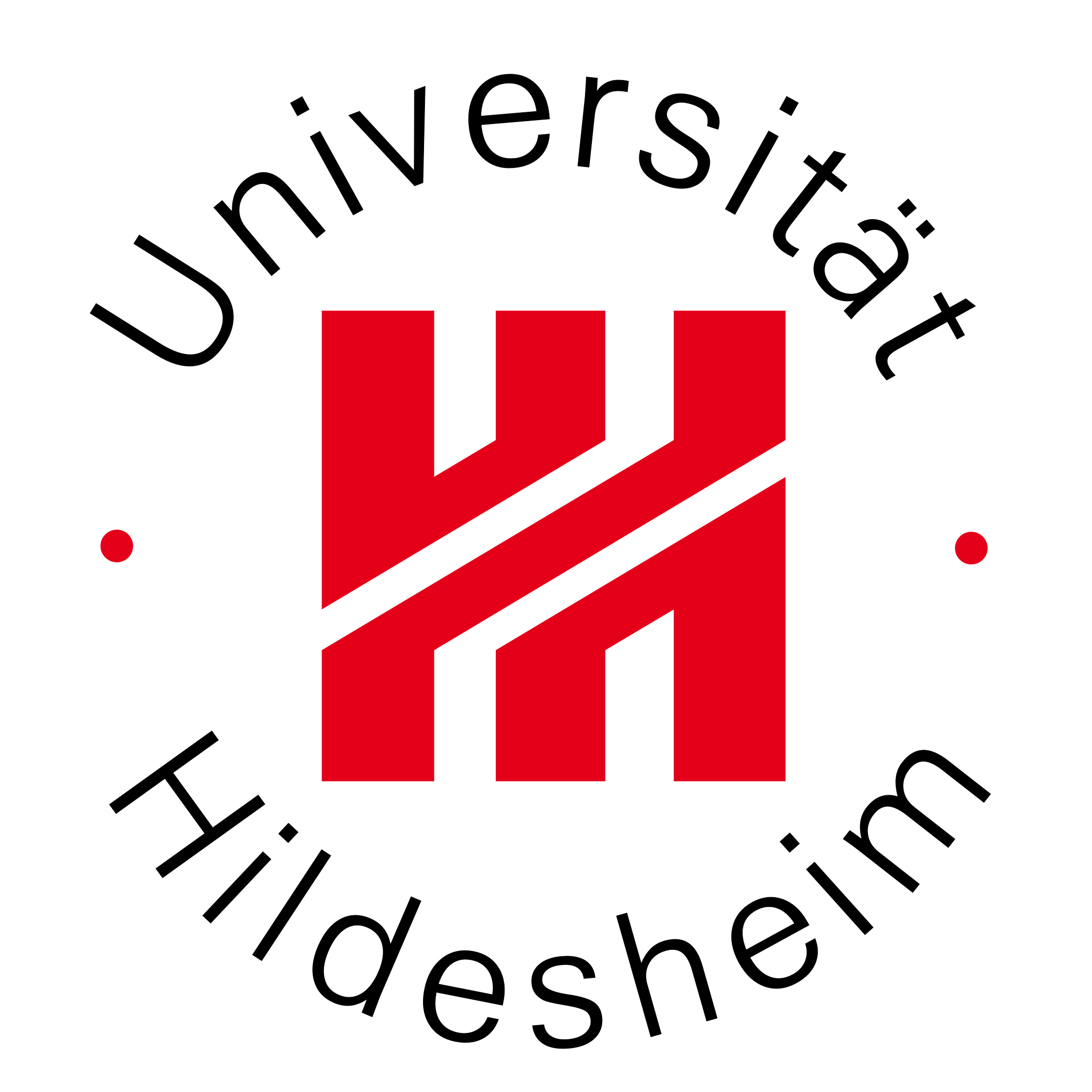}
\vspace{2cm}


{ \Large \bfseries Global Context Enhanced Anomaly Detection of Cyber Attacks via Decoupled Graph Neural Networks}\\  
\vspace{0.5cm}
\vspace{1.5cm}

\begin{minipage}{0.46\textwidth}
\begin{flushleft} \large
\emph{Author:}\\
Ahmad \textsc{Hafez} \\ 
302968 \\ 
\end{flushleft}
\end{minipage}
\begin{minipage}{0.49\textwidth}
\begin{flushright} \large
\emph{Supervisors:}\\
\mbox{Prof. Dr. Dr. Lars \textsc{Schmidt-Thieme}}\hfill \\ 
Nourhan \textsc{Ahmed} 
\end{flushright}
\end{minipage}\\


\vspace{2cm}


{ 19th September 2022 }\\ 
\vspace{1cm}

{ \large \bfseries Thesis submited for}\\  
\vspace{0.3cm}
\textsc{\Large Master of Science in Data Analytics}\\
\vspace{1cm}
\textsc{\large Wirtschaftsinformatik und Maschinelles Lernen}\\ 
\vspace{0.5cm}
\textsc{\large Stiftung Universität Hildesheim}\\ 
\vspace{0.3cm}
\textsc{\large Universitatsplätz 1, 31141 Hildesheim}\\ 

\vfill 

\end{titlepage}

\setcounter{secnumdepth}{1}


\noindent \textbf{Statement as to the sole authorship of the thesis:}
\vspace{0.4cm}
\\Global Context Enhanced Anomaly Detection of Cyber Attacks via Decoupled Graph Neural Networks.
\\I hereby certify that the master's thesis named above was solely written by me and that no assistance was used other than that cited. The passages in this thesis that were taken verbatim or with the same sense as that of other works have been identified in each individual case by the citation of the source or the origin, including the secondary sources used. This also applies for drawings. sketches, illustration as well as internet sources and other collections of electronic texts or data, etc. The submitted thesis has not been previously used for the fulfilment of a degree requirements and has not been published in English or any other language. I am aware of the fact that false declarations will be treated as fraud.
\vspace{7cm}

{ 19th September 2022}, Hildesheim

\thispagestyle{empty}
\setcounter{tocdepth}{2}
\newpage


\begin{abstract}

Recently, there has been a substantial amount of interest in GNN-based anomaly detection. Existing efforts have focused on simultaneously mastering the node representations and the classifier necessary for identifying abnormalities with relatively shallow models to create an embedding. Therefore, the existing state-of-the-art models are incapable of capturing nonlinear network information and producing suboptimal outcomes. In this thesis, we deploy decoupled GNNs to overcome this issue. Specifically, we decouple the essential node representations and classifier for detecting anomalies. In addition, for node representation learning, we develop a GNN architecture with two modules for aggregating node feature information to produce the final node embedding. Finally, we conduct empirical experiments to verify the effectiveness of our proposed approach. The findings demonstrate that decoupled training along with the global context enhanced representation of the nodes is superior to the state-of-the-art models  in terms of AUC  and introduces a novel way of capturing the node information.


\vfill
\end{abstract}

\newpage

\tableofcontents
\listoffigures
\listoftables
\newpage
\pagenumbering{arabic}

\chapter{Introduction}
The purpose of cybersecurity, information technology security (IT security), or computer security is to protect computer systems and networks from the disclosure of information, theft of, or damage to their hardware, software, or electronic data, as well as the disruption or misdirection of the services that they provide\cite{schatz2017towards}.

As a result of an increased dependence on computer systems, the Internet, and wireless network standards such as Bluetooth and Wi-Fi, as well as the proliferation of 'smart' devices such as smartphones, televisions, and other devices that make up the Internet of Things (IoT), the field of cybersecurity has become more important \cite{stevens2018global,kianpour2021systematically}.

Anomaly is a noun, that can be defined as something or a person that is different from what is usual, or unusual enough to be noticeable \cite{dictionary1989oxford,staff2004merriam,everitt2010cambridge,ahmed2016survey}
According to \cite{chandola2009anomaly} -the most cited (survey) article on anomaly detection- Anomaly detection is defined as the challenge of identifying data patterns that deviate from typical behavior.
Depending on the research field and researchers background, they might use different terms or synonyms for anomalies like: nonconforming patterns, outliers, discordant observations, exceptions, aberrations, surprises, peculiarities, or contaminants.

Anomaly detection has several challenges that can be summed in the following list:
\begin{itemize}
  \item The borderline between normal and abnormal activity is usually not clear.
  \item If the anomaly is associated with a cyber attack, the attackers adjust their actions to look normal, resulting in a harder identification of such attacks.
  \item For some fields and due to the fluidity, what is considered as a borderline between normal and abnormal can change with time.
  \item Anomaly detection can be considered to be very field-specific, i.e. an anomaly detection methodology for a certain field most probably is not usable for other fields.
  \item The lack of labeled data for supervised or semi-supervised model training.
  \item Noisy data can result in harder identifications of anomalies.
  \end{itemize}
There are mainly three different types of anomalies, that can be defined as follows:
\begin{itemize}
\item Point anomalies: This kind of anomaly appears when a single point in the data can be regarded an anomaly with respect to other points in the data as shown in figure \ref{fig:panom}.
\begin{figure}[!ht]
    \centering
    \includegraphics[scale=0.3]{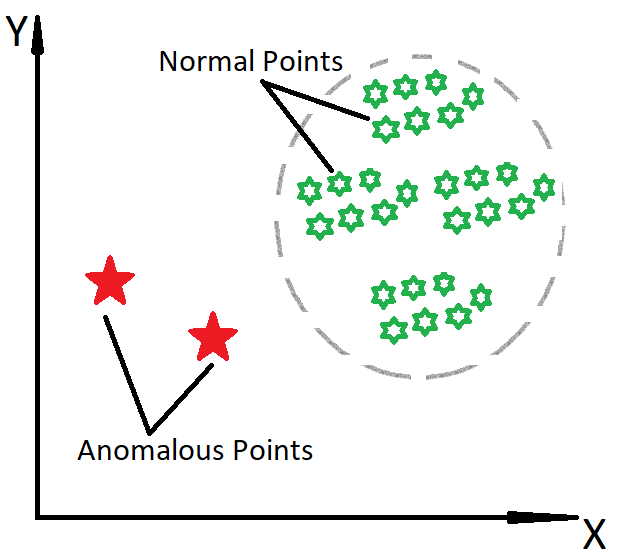}
    \caption{Example of point anomalies}
    \label{fig:panom}
\end{figure}
\item Conditional/Contextual Anomalies: This type of anomaly occurs when the data sample is anomalous with respect to some context.
\item Collective Anomalies: This class of anomalies emerges when a collection or subgroup of data points or instances is regarded anomalous, but individual data instances are not considered.
\end{itemize}  
Different Methods of Anomaly Detection:
\begin{itemize}
\item Supervised Anomaly Detection: As the name supervised suggests, this expects the existence of a labeled training dataset for 2 classes normal and abnormal or anomaly which can be used to train a biclassifier, unfortunately most of the available datasets are not labeled and, as the anomaly name suggests, the datasets are imbalanced, i.e. the normal points outnumber the abnormal points.
\item Unsupervised Anomaly Detection: Again, as the term unsupervised suggests the data here is not labeled, so the algorithms depends on the hypothesis that the normal instances outnumber the abnormal (anomaly) ones, if this hypothesis fails then a high false discovery rate is encountered.
\item Semi-supervised Anomaly Detection: This kind of algorithms uses datasets with labeled normal instances.\cite{villa2021semi}
\end{itemize}




\section{Motivation}
The smart grid topic from my point of view is a sweet spot between different electrical engineering and computer science topics, and due to my interest in such interdisciplinary topics, I wanted to revisit smart grids, but this time from a new perspective and aspect, which was cybersecurity issues.

Lately, I noticed a growing interest in smart grid security, and multi million dollar funds for such research projects funded by the American department of energy\cite{DOE}, trying to understand the vulnerabilities and using the machine learning to identify the anomalies would be an interesting topic to write a thesis on.

Adding to the previous reasons, the costs of cyber attacks are increasing dramatically according to the cybercrime magazine ``Cybersecurity Ventures'': The annual costs of cyber crimes are expected to reach 10,5 trillion dollars by 2025\cite{morgan_2021}.
According to IBM's projections, the 2022 mean costs of a data breach is 4.35 million US dollars, which is 2.6\% increase from 4.24 million US dollars in 2021 \cite{IBM}.
According to Forbes, there are daily more than 26 thousand cyberattacks\cite{Forbes}.

Graph Neural Networks (GNN) can analyze patterns and learn from them which is applied to cybersecurity systems to help preventing similar attacks and respond to them as the graph takes into consideration the structure of the network as the graph contains the nodes and the edges show how different entities or nodes are connected to each other, also how strong the bond between these different entities is shown with the weighted edges, and in case the connection is directed or undirected. Most of the current GNN-based cybersecurity models use unsupervised learning for anomaly detection, in this work classification -which is considered as a supervised learning problem- is utilized because the supervised learning results are more accurate and reliable compared to unsupervised learning techniques due to the fact that supervised learning depends on ground truth not just getting the relationships between the dataset different points without guidance or a right answer.

\section{Problem Statement}

The vast majority of the datasets that are available do not have labels; however, the tasks themselves are supervised learning ones. Furthermore, the vast majority of the work that has been done on cyber attack detection uses unsupervised anomaly detection. This is in addition to the reality that the datasets are highly imbalanced, due to the fact that the definition of anomalies, as well as networking behavior, differs between parties that are acting on various scales, domains, cultures, and geolocations.

More formally, let's define our network problem as a graph

$G=\left(V, A, X, Y^{L}\right)$,
$V$ is the set of nodes, where nodes represent users and objects, but only the identification of abnormal or unusual users is of concern.
,
$A$ is the adjacency matrix of the nodes, which is used to map the associations between the different nodes in the network, where edges or links describe users behaviours.
$X$ is the initial feature matrix
$Y^{L}$ are the partial labels on nodes 1 for abnormal and 0 for normal Predictive Function $F: G \rightarrow Y$
$Y=Y^{L} \cup Y^{U}$ where $Y^{U}$ unobserved labels of the nodes in $G$

The Predictive function $F$ is divided into:
\begin{enumerate}
    \item GNN encoder $g: V \rightarrow R^d$ Encodes the structure patterns into the node representations \\ 
    $H=g(G, \theta)$
    \item
Binary classifier $f: R^d \rightarrow\{0,1\}$
\\ In most cases, it is carried out on top of the output of node representations by $g$ to distinguish the normal and abnormal nodes.
\\$\hat{Y}=f(H, \phi)$
    
\end{enumerate}
\section{Research Procedure}
The purpose of this study was to investigate the various effects that varying the building blocks of the decoupled training proposed by Yanling Wang et al. in their paper Decoupling Representation Learning and Classification for GNN-based Anomaly Detection \cite{wang2021decoupling} would have. The model, in its most basic form, is divided into two stages: a pre-training stage and a tuning stage. The pre-training stage uses the graph data as an input, and the GNN encoder and self-supervised loss that the GNN learns based on it. The tuning stage uses the GNN encoder and self-supervised loss that the GNN learn

For instance, the encoder for the GNN that was proposed was a Graph Isomorphism Network; however, a simple yet beneficial improvement would have been to employ a multi-encoder rather than the pure Graph Isomorphism Network.
Other modifications were implemented, but none of them led to an improvement that was either significantly better or significantly larger.
\section{Contributions}
This study's primary contribution was to combine the utilization of more than one graph encoder and averaging their outputs for representation learning \cite{leng2021enhance}, as well as decoupling the representation learning and the classification \cite{wang2021decoupling}, in order to achieve better AUC results on five distinct datasets, namely Wiki, Reddit, Bitcoin Alpha, and Amazon, as well as CIC-IDS2017.

\section{Outline}
This thesis is organized as follows:
\\
Theory \textbf{(Chapter \ref{ch2})}: This chapter provides an overview of the fundamental concepts underlying this research, including artificial intelligence, machine learning, deep learning, representational learning, and GNNs, beginning with their respective definitions and short introductions about their development.
\\
Related Work \textbf{(Chapter \ref{ch3})}: This chapter provides a literature review of the current state-of-the-art in the field of graph based anomaly detection regardless of whether supervised or semi-supervised.
\\
Methodology \textbf{(Chapter \ref{ch4})}: In addition to the definitions, the problem formulation, and the design, this chapter delves into the baseline methodology.
\\
Experiments \textbf{(Chapter \ref{ch5})}: In this chapter, the various datasets that were used to train the models are presented, along with the preprocessing steps that were carried out on them in order to get them ready for the training and gauging of the various models.
\\
Conclusion \textbf{(Chapter \ref{ch6})}: This chapter presents both a review of the results that were obtained throughout the course of this study as well as a look ahead to potential future work.
\chapter{Theory}
\label{ch2}
In this chapter, a gentle introduction to Machine Learning (ML), Deep Learning (DL), Representation Learning (not RL as it is reserved for reinforcement learning) and with a bit detailed Graph Neural Networks (GNN) as Deep learning is a type of representation learning and the representation learning is a type of the machine learning and the machine learning.

\section{Machine Learning}
"Machine Learning" term was coined by Arthur Samuel in 1959 \cite{samuel1988some}.
Machine learning is considered the most important AI tool and is becoming more and more reliable day after day due to the big data generated.

There are mainly four types of machine learning algorithms or paradigms, \nameref{SL}, \nameref{USL}, \nameref{RL} and \nameref{SSL} \cite{mohammed2016machine,ayodele2010types}.
\subsection{Supervised Learning}\label{SL}
Supervised machine learning is an input-output mapping.
In supervised learning, we have a dataset of inputs and labeled or ground truth correct outputs, so the model can learn from this data either through \nameref{class} or \nameref{regression}.

\paragraph{Classification}\label{class}: is used, as one could infer from the name, to classify the test points into certain groups i.e. it is used to predict discrete target values. It does this by identifying particular entities within the dataset and attempting to draw some inferences about how those entities ought to be labelled or recognized. Linear classifiers, Support Vector Machines (SVMs), decision trees, K-Nearest Neighbor(KNN or k-NN), and random forest are some of the most common classification algorithms\cite{sarker2021machine}.
\paragraph{Regression}\label{regression}:
is utilized in order to determine how the dependent variable and the independent ones interact with one another. It is utilized for the purpose of estimating, predicting, and projecting continuous amounts and targets, such as for the amount of income a corporation generates through sales. The linear regression, the logistical regression, and the polynomial regression are three well-known types of regression algorithms\cite{sarker2021machine}.
\subsection{Unsupervised Learning}\label{USL}
Unsupervised learning makes use of data that hasn't been tagged, i.e. unsupervised learning algorithms train the models using unlabeled data, in other words, its goal is to find out the relationships between data points and get insights about the data itself not predicting outcome of a new data point. It extracts patterns from the data and uses them to solve clustering and association problems. When subject matter experts are unsure of common properties within a data set, this is especially valuable. Hierarchical, k-means, and Gaussian mixture models are common clustering algorithms\cite{sarker2021machine}.
\subsection{Reinforcement Learning}\label{RL}
Reinforcement learning is a form of machine learning techniques that allows software agents and computers to automatically assess the optimal behavior in a specific environment or context to improve its efficiency. This type of learning takes an environment-driven approach, which can be thought of as an environment-driven approach. This form of learning is based on receiving a reward or being subjected to a penalty, and its ultimate objective is to use the insights received from environmental activists to take action that will either raise the reward or decrease the penalty. It is not recommended to use it for the purpose of resolving simple or uncomplicated issues because it is an useful tool for training ML models that can boost automation or optimize the operational efficiency of sophisticated systems such as robotic systems, autonomous driving tasks, manufacturing, and supply chain logistics. However, it can help improve automation or optimize the operational efficiency of these systems\cite{sarker2021machine,kaelbling1996reinforcement,mohammed2016machine}.

\subsection{Semi-supervised Learning}\label{SSL}
Semi-supervised learning is not always considered one of the main paradigms or algorithms of machine learning and is attempts at merging the supervised learning and the unsupervised learning,i.e. semi-supervised learning is used when just a subset of the input data will have labels attached to it, so unsupervised learning will be used to generate pseudolabels i.e. 
the unlabeled part will be used with unsupervised learning, the labeled part of data will be suitable for supervised learning and \cite{van2020survey,sarker2021machine,seeger2000learning,chawla2005learning,zhu2005semi,zhu2009introduction,ericsson2022self}.

\section{Deep Learning}\label{DL}
The deep learning has several names due to historical reasons, for example it was called cybernetics between the forties and sixties of the twentieth century, then it was referred to as connectionism and artificial neural networks in the eighties and nineties of the previous century, then the name deep learning took over by the beginning of the second half of the first decade of the current century.
The current expression "deep learning" gives an impression that the field is currently wider than the neurally influenced frameworks\cite{goodfellow2016deep}.
Deep neural networks as a research topic is a subset of machine learning that is more than a 2-layer neural network. These neural networks "learn" from big data. Compared to machine learning, deep learning automates feature extraction/engineering. \cite{russell2010artificial,han2022data}.


\subsection{Artificial Neural Networks}

In this section, the feedforward neural networks or Multilayer perceptrons topic is introduced, starting with the Artificial neural networks (ANNs) -as the name suggests- are influenced by biological systems, particularly the functioning of brain neurons.
There are around 10 billion neurons in the human brain, and each one is linked to almost 10 thousand other neurons. With this description of the human brain it can be considered as a sophisticated nonlinear system made up of neurons that are capable of carrying out complicated tasks\cite{goodfellow2016deep,haykin2009neural}.

The history of the artificial networks goes back to the first half of the twentieth century, An article on the potential function of neurons was written in 1943 by neurophysiologist Warren McCulloch and mathematician Walter Pitts\cite{mcculloch1943logical}.
The McCulloch-Pitts model of a neuron serves as the foundation for Rosenblatt's perceptron\cite{rosenblatt1958perceptron}.

A simple but important question would be how the Artificial Neural Networks work?, to answer this question let's go back to the name one more time, the neural network is mimicking how the brain functions as the network learns from its surroundings and weights are attached to inter-neuron connections to keep the learned information.

The neural model consist mainly of three fundamental components\cite{haykin2009neural}:
\begin{enumerate}
   \item A group of connecting links, or synapses: each of the connecting links has a distinct weight or strength that can even be a negative or positive number. 
   \item Adder: for aggregating the input signals scaled by the various synaptic strengths of the neuron.
   \item Squashing or Activation Function: is a function that squashes or restricts neuron's output amplitude to a limited extent which means that activation functions are adding a non-linearity to the model.
\end{enumerate}

\begin{figure}[htp]
    \centering
    \includegraphics[width=\textwidth]{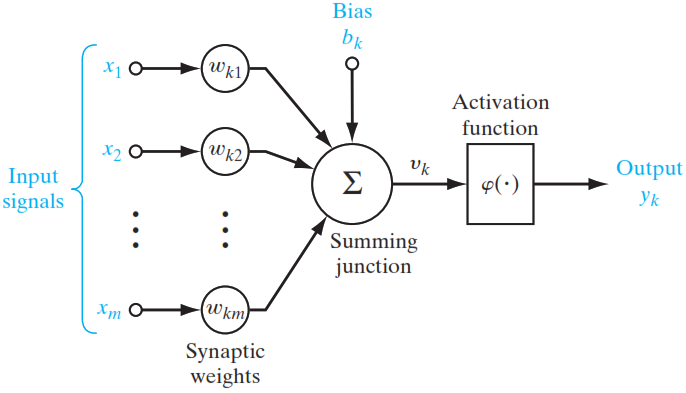}
    \caption{Nonlinear model of a neuron {k}.}\cite{haykin2009neural}
    \label{fig:nk}
\end{figure}

For neuron \emph{k} shown in \autoref{fig:nk} the following steps are taking place:
\begin{enumerate}
   \item The input signals $ x_1, x_2, \ldots, x_m $ are multiplied by the corresponding weights $w_{k1},w_{k2}, \ldots, w_{km}$
\begin{equation}
u_{k}=\sum_{j=1}^{m} w_{k j} x_{j}
\end{equation}
\item Then the biases $b_k$ are added, getting $v_k$ as the output of the adder.
\begin{equation}
v_{k}=u_{k}+b_{k}
\end{equation}
\item Finally, the activation function $\varphi(.)$ is applied to add non-linearity to the model, having $y_k$ as the output of the neuron.
\begin{equation}
y_{k}=\varphi\left(v_{k}\right)
\end{equation}
\end{enumerate}

As a conclusion, Multilayer perceptrons (MLPs) work as a function estimator, as the input of each perceptron is a function of the output of the previous perceptron.
MLPs are used with tabular data for classification or regression problems.
\subsection{Convolutional Neural Networks}\label{cnn}
The Convolutional Neural Networks (ConvNets  or CNNs) are used for image data or the data has spatial relationships, CNNs are also used for classification or regression problems.
Through the work that Kunihiko Fukushima and Yann LeCun did in 1980 \cite{fukushima1980self} and 1989 \cite{lecun1989backpropagation}, respectively, they set the groundwork for the field of research that is now known as convolutional neural networks. More notably, Yann LeCun was able to train neural networks to locate and recognize patterns within a set of handwritten zip codes by effectively applying backpropagation. Throughout the 1990s, he and his colleagues would continue their study, culminating in the development of "LeNet-5" \cite{lecun1989backpropagation,lecun1989generalization,lecun1989handwritten,lecun1995convolutional,lecun1998gradient} which applied the similar concepts discovered in earlier research to the process of document recognition \cite{lecun1998gradient}. Since then, a variety of other CNN architectures have come into existence as a result of the release of new datasets, such as MNIST \cite{lecun1998mnist} and CIFAR-10, and the hosting of contests, such as the ImageNet Large Scale Visual Recognition Challenge (ILSVRC) \cite{ibmcloudeducationcnn_2020}.
A discrete convolution over 2D image I, and a 2D Kernel K can be written as follows\cite{goodfellow2016deep}:
\begin{equation}
S(i, j)=(K * I)(i, j)=\sum_m \sum_n I(i-m, j-n) K(m, n)
\end{equation}
\subsubsection{CNN architecture}
The three primary types of layers that CNNs use: Convolutional layer, Pooling layer, Layer that is Fully Connected (FC).
\paragraph{Convolutional layer}
A convolutional network will often begin with a convolutional layer as its initial layer, During the process of scanning the input I with regard to its dimensions, the convolution layer (CONV) makes use of filters that are able to carry out operations that include convolution. The filter size F and the stride S are both included in its hyperparameters. The output of the convolutional layer (O) is known as a feature map or activation map depending on its purpose.

For a convolutional layer with input volume size W, S stride, P zero padding and size of its layer neurons F then number of neurons N (output size of feature map) can be calculated using the following equation\cite{goodfellow2016deep}:
\begin{equation}
N=\frac{W-F+2 P}{S}+1
\end{equation}
convolutional layers can be followed by either more convolutional layers or pooling layers\cite{goodfellow2016deep}
\paragraph{Pooling layer}
A pooling layer, abbreviated as "POOL," is a downsampling process that is commonly used after a convolution layer. This layer achieves some degree of spatial invariance. Pooling can be broken down into several subcategories, the most notable of which are the maximum pooling and the average pooling. Both of these types of pooling take the respective maximum and average values into consideration, maximum pooling is used more than the average pooling due to the fact that average pooling downsamples the feature map but maximum pooling maintains the features identified.
\paragraph{Fully-connected layer}
The fully-connected layer is the final layer, and it is utilized in order to achieve optimal results like class scores, although The intricacy of the CNN grows with each additional layer, enabling it to recognize an ever-growing percentage of the image. The earlier layers concentrate on the more fundamental characteristics, such as colors and borders. As the image data is passed through the layers of the CNN, it first begins to detect the object's major components or forms and continues to do so until it finally recognizes the object for which it was designed.
\begin{figure}[H]
    \centering
    \includegraphics[width=\textwidth]{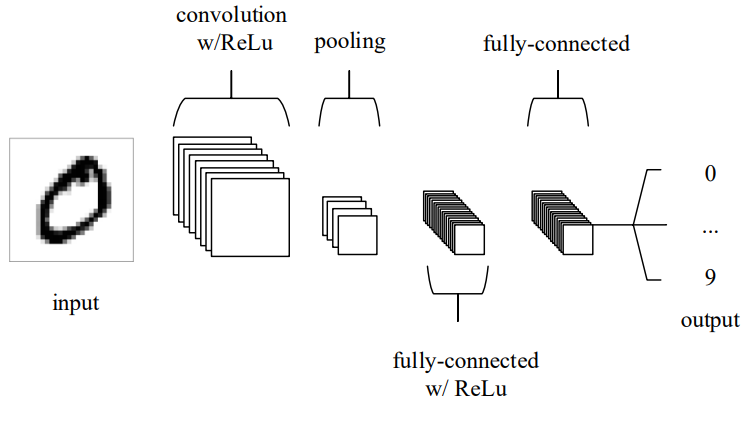}
    \caption{Basic CNN architecture to recognise a hand-written decimal number}\cite{o2015introduction}
    \label{fig:cnn}
\end{figure}
A feature map that summarizes the existence of features that have been recognized in the input is produced by convolutional neural networks by applying a filter to an input.
The filter or kernel has a set of weights to be multiplied by the input and operates as a sliding window across the entire image, allowing CNNs to pick up details from nearby cells.
Convolutional neural networks' unique feature is its ability to learn the filters while being trained on a particular prediction problem.

\subsubsection{CNN loss functions}
Different loss functions can be used with CNNs for example: Contrastive loss, Hinge loss,  Triplet loss, Softmax loss, Kullback-Leibler Divergence\cite{gu2018recent}.

\subsection{Recurrent Neural Networks}\label{RNN}
Recurrent Neural Networks (RNNs) are used for sequence data like text or speech regression and classification problems, Long short-term memory (LSTM) \cite{hochreiter1997long} is one of the most common RNNs, LSTMs have feedback connection which sets it apart from feedforward neural networks \cite{salehinejad2017recent}.

\subsubsection{Types of RNNs and their applications}
\begin{enumerate}
   \item One-to-one: it is a kind of RNN that gives one output and given one input, it is the basic RNN and can be used for image classification.
   \item One-to-many: it is a kind of RNN that gives more than one output and given one input, it can be used for image captioning and music generation.
   \item Many-to-one: it is a kind of RNN that gives one output and given more than one input, it can be used for sentiment analysis.
   \item Many-to-many: it is a kind of RNN that gives more than one output and given more than one inputs, it can be used for machine translation and number of inputs wont be exactly as the number of outputs, and can be used too for name-entity recognition but in this case the number of inputs and outputs must be equal.
   \end{enumerate}
   
RNNS suffer from a exploding and vanishing gradients, this proble arises due to ther fact that multiplication of gradients either increase causing the exploding gradient or decrease causing the vanishing gradient, the solution of exploding gradient can be gradient clipping and for vanising gradient is the usage of LSTM.
\subsubsection{RNN loss functions}
The all time RNN loss function is a summation of the loss at each time step as shown below:
\begin{equation}
\mathcal{L}(\widehat{y}, y)=\sum_{t=1}^{T_y} \mathcal{L}\left(\widehat{y}^{<t>}, y^{<t>}\right)
\end{equation}
\subsection{Activation Functions}
As above mentioned the activation function is used to limit the output adding non linear complexity to the models, the choice of the activation function for hidden layers depends on the network type; Multilayer Perceptron (MLPs), Convolutional Neural Networks (CNNs) or Recurrent Neural Networks (RNN), on the other hand the activation function for output layer depends on the problem whether binary classification, multiclass classification or regression.
Most common activation functions are: Sigmoid, Tanh, Rectified linear units (ReLU), Softmax and Linear.
\subsubsection{Sigmoid function}
Sigmoid function output or range is limited between [-1,1], and is used as the hidden layers activation function for Recurrent Neural Networks (RNN), also the sigmoid is used as the output layer activation function in case of binary or multilabel classification.

The sigmoid function is defined as follows:
\begin{equation}
\operatorname{sig}(x)=\frac{1}{1+e^{-x}}=\frac{e^{x}}{1+e^{x}}=\frac{1}{2} \cdot\left(1+\tanh \frac{x}{2}\right)
\label{eqn:simoid}
\end{equation}

\subsubsection{hyperbolic tangent (tanh) function}
Tanh function output or range is: [-1,1], and is used as the hidden layers activation function for recurrent Neural Networks (RNNs).
\begin{equation}
tanh(x)=\frac{e^{x}-e^{-x}}{e^{x}+e^{-x}}
\end{equation}
\subsubsection{Rectified Linear Unit (ReLU) function}
ReLU activation functions are used for hidden layers of Multilayer Perceptron (MLPs) and Convolutional Neural Networks (CNNs).
\begin{equation}
ReLU(x)=x^+=\max (0, x)
\end{equation}
\subsubsection{Leaky Rectified Linear Unit (Leaky ReLU) function}
Leaky ReLU activation function is built on a ReLU, but it's different from Relu as instead of returning zero for negative values it has a little slope for negative values, and the slope is a hyper-parameter as it is not learnt by training.
\begin{equation}
ReLU(x)=\max (\epsilon x, x)
\\with \epsilon<<1
\end{equation}

\subsubsection{Exponential Linear Unit (ELU)}
A modification of the ReLU function that is differentiable at any value eliminates vanishing gradient problem
\begin{equation}
ELU(x)=\max (\epsilon(e^x-1), x)
\\with \epsilon<<1
\end{equation}
\subsubsection{Softmax function}
Softmax output is a probability, so its range is [0,1] and is used as the output layer activation function in case of multiclass classification.
\begin{equation}
\sigma(\vec{z})_{i}=\frac{e^{z_{i}}}{\sum_{j=1}^{K} e^{z_{j}}}
\end{equation}

\subsubsection{Linear function}
When the activation is proportionate to the input, the activation function is said to be linear. This function is also sometimes referred to as "no activation" or the ``identity function'' (when multiplied by 1.0).

The function does not alter the weighted total of the input in any way; rather, it only returns the value that it was provided with.

\section{Representation Learning}
Information representation determines to some extent the easiness of a learning task, hidden layers of a feedforward artificial neural network can be considered as a representation to the output layer as the hidden layers main task is to learn to support the last output layer with a representation this might be a reason that representation learning is called also feature learning, in other words representation learning tasks are intermediate tasks with goal of easing learning and a better representation is one that eases the following learning; in this work there is a concentration on semi-supervised learning which is eased with representation learning, as in most of real life cases training datasets have very little labeled data compared to unlabeled data, so representation learning will help learning from unlabeled data and labeled data resulting in less overfitting problems arise with the usage of only labeled data which are less in the amount for supervised learning, for the unlabeled data, we can acquire effective representations, and we can utilize these representations to complete the supervised learning challenge.\cite{goodfellow2016deep,bengio2013representation}.
It is worth mentioning that latent space or embedding space importance comes from the fact that there is always some kind of hidden features -latent itself means hidden- so through embedding in latent space where similar items will be closer to each other so computers can computationally ``understand'' relationships/similarities between items topping one-hot encoding as one-hot encoding does not show similarity measures; as an example of the embedding techniques can be the word2vec developed in 2013 by Tomas Mikolov \cite{mikolov2013distributed,mikolov2013efficient}, Word2vec receives as input a huge corpus of text and outputs a vector space, which typically has several hundred dimensions. Within this vector space, a matching vector is assigned to each unique word that was included in the input corpus. The placement of word vectors in the vector space is done in such a way that words that are found in the corpus in similar contexts are clustered together in close proximity to one another in the space.

\section{Graph Neural Networks}
\begin{figure}[!ht]
    \centering
    \includegraphics[width=\textwidth]{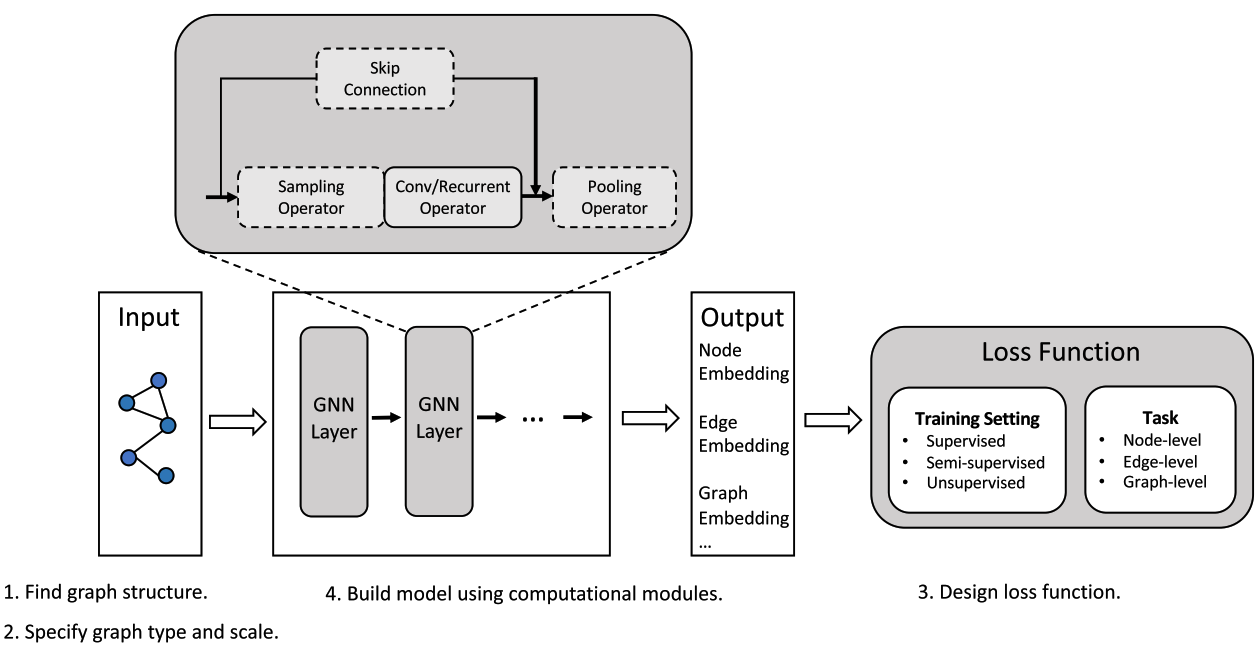}
    \caption{A pipeline for a GNN model} \cite{zhou2020graph}
    \label{fig:gnn}
\end{figure}

This part is heavily based on the survey and literature review work of \cite{wu2020comprehensive,zhou2020graph}. First point to discuss would be what is the reason behind using graphs, and a simple, yet realistic reason will be because graphs are literally everywhere, graphs can describe different relations like sequences or grids. A generic design process for a GNN model is shown in \autoref{fig:gnn} \cite{zhou2020graph}.

\subsection{Evolution of graph neural networks}
The first application of graph neural networks can be traced to the end of the tweentith century percisely, 1997 when Sperduti and Starita applied the Recursive Neural Networks to directed acyclic graphs\cite{sperduti1997supervised}. In 2005, Gori et al.  coined the term "Graph Neural Network" in their paper \cite{gori2005new}. These early researches are considered to be examples of recurrent graph neural networks(RecGNNs) as they learn node's representation using neighbors information resulting in a computationally expensive process.
Influenced by the high performance of convolutional neural networks, the first Convolutional Graph Neural Network (ConvGNN) was the spatial-based ConvGNN by Micheli et al. \cite{micheli2009neural} published in 2009, the spectral-based ConvGNNs appeared almost 5 years later as the firs spectral-based ConvGNN was published in 2014 by Bruna et al \cite{bruna2014spectral}.

\subsection{Taxonomy of Graphs}
The graphs might be classified based on three different dimensions first dimension would be the Direction i.e. whether the graph is directed or not second would be Homogeneity whether the graph nodes are of same type or not and same goes for edges and third and last dimension is Time variance, whether the graph is static not changing with time or dynamic that changes with time.
\subsubsection{Directed/Undirected Graphs}
Directed graphs, as contrast to undirected graphs, have all of their edges going in a direction from one node to another in order to convey more information. Each edge in a graph that is not directed can alternatively be thought of as two directed edges in some cases.
\subsubsection{Homogeneous/Heterogeneous Graphs}
In homogeneous graphs, the kinds of nodes and edges are identical, whereas in heterogeneous graphs, the types of nodes and edges are distinct from one another. It is vital to give more thought to the types of nodes and edges used in heterogeneous networks because of the important functions they play in these graphs.

\subsubsection{Static/Dynamic Graphs}
A graph is considered to be dynamic when either the features that are input or the topology of the graph change over the course of time. In dynamic graphs, it is important to take the temporal information into careful consideration\cite{hafez2021convdysat,zhou2020graph}.
\subsection{Graph Learning Task Types}
There are three distinct levels or types of graph learning tasks, and these are the Node-level, the Edge-level and the Graph-level respectively.
\subsubsection{Node-level}
Node-level tasks are those that concentrate on nodes and might include things like node classification, node regression, and node clustering, amongst other things. The process of node classification endeavors to group nodes into a number of distinct groups, while the process of node regression forecasts a continuous value for each node. The objective of node clustering is to divide the nodes into a number of distinct groups, with the expectation that nodes with similar properties will be grouped together. By use of information propagation and/or graph convolution, RecGNNs and ConvGNNs are capable of extracting high-level node representations. GNNs are able to do node-level tasks in an end-to-end manner if the output layer is either a multi-perceptron or a softmax layer.

\subsubsection{Edge-level}
Edge-level tasks include edge classification and link prediction. These tasks require the model to either categorize the different types of edges or predict whether or not there is an edge connecting two given nodes. Predicting the label or connection strength of an edge can be accomplished with the help of a neural network or a similarity function if the hidden representations of two nodes obtained from GNNs are used as inputs.
\subsubsection{Graph-level}
Graph-level tasks come in the form of graph classification, graph regression, and graph matching; in order to complete any of these tasks, the model must first learn graph representations. GNNs are frequently used with pooling and readout procedures in order to achieve the goal of obtaining a compact representation on the graph level\cite{hamilton2020graph,wu2020comprehensive}.
\subsection{Graph Learning Training Settings}
Graph learning tasks can also be broken down into three distinct types of training scenarios when viewed from the perspective of supervision.
\subsubsection{Supervised}
A labeled data collection is made available for training purposes when supervised conditions are used, and it is very common with graph-level classification \cite{wu2020comprehensive}, the following four algorithms are examples of supervised graph classification models: \cite{zhang2018end,ying2018hierarchical,pan2015joint,pan2016task}, in the work of Zhang et al. \cite{zhang2018end}, the authors propose a Deep Graph Convolutional Neural Network (DGCNN) that takes as an input a set of labeled graphs and classifies them, as shown in figure \ref{fig:ggcn}.
\subsubsection{Semi-supervised}
A semi-supervised learning environment provides a limited number of labeled nodes for training in addition to a substantial number of unlabeled nodes. During the test phase, the transductive option demands the model to make predictions about the labels of the unlabeled nodes that are provided. On the other hand, the inductive setting gives the model additional unlabeled nodes to infer that come from the same distribution. The majority of tasks involving node and edge classification are semi-supervised.\cite{wang2020unifying,rossi2018inductive} are the most recent researchers to implement a combined transductive-inductive scheme, with the intention of paving a new road towards a mixed setting.
\subsubsection{Unsupervised}
In an unsupervised situation, the data that the model uses to discover patterns are always unlabeled. The task of node clustering is a common example of unsupervised learning, and it is very common with graph embedding tasks \cite{wu2020comprehensive} making use of an autoencoder to embed into a latent representation the graph with an encoder using graph convolutional layers and a decoder to rebuild the graph structure as example of literature under this area: \cite{kipf2016variational, pan2018adversarially}, moreover in the work of Kipf and Willing a variational graph autoencoder was introduced to learn undirected graphs latent representations\cite{kipf2016variational}.

\subsection{Classification of Graph Neural Networks}
This section is dedicated to presenting \cite{wu2020comprehensive} taxonomy of GNNs. According to \cite{wu2020comprehensive}, GNNs can be broken down into four distinct categories: recurrent graph neural networks (RecGNNs), convolutional graph neural networks (ConvGNNs), graph autoencoders (GAEs), and spatial-temporal graph neural networks (STGNNs). In the following, we will provide an overview of each category in brief form.
\subsubsection{Recurrent graph neural networks}
The majority of recurrent graph neural networks(RecGNNs), are considered to be pioneer works of graph neural networks. The goal of RecGNNs, which use recurrent neural architectures, is to learn node representations.
They work under the assumption that each node in a network engages in a continuous information or message exchange with the nodes that surround it until a stable equilibrium is achieved. RecGNNs are significant conceptually and served as an inspiration for subsequent research on convolutional graph neural networks. In specifically, the concept of message passing is something that is passed down across the generations in spatial-based convolutional graph neural networks\cite{scarselli2008graph, gallicchio2010graph,li2015gated,dai2018learning}.

\subsubsection{Convolutional graph neural networks}
Convolutional graph neural networks (ConvGNNs) are a type of graph neural networks that generalize the convolutional operation by moving it from grid data to graph data. The primary objective is to construct a representation of a node by combining the properties of both the node itself and its neighbors. ConvGNNs, which are not the same as RecGNNs, are built by stacking numerous graph convolutional layers in order to obtain high-level node representations. ConvGNNs are critical components that must be included when developing a wide variety of more complicated GNN models.

ConvGNNs can be broken down into two distinct categories, namely spectral and spatial models, putting the Spectral-based ConvGNNs and Spatial-based ConvGNNs a side by side for comparison The theoretical underpinnings of spectral models can be found in graph signal processing.

One way to construct new ConvGNNs is to develop new graph signal filters, such as Cayleynets \cite{levie2018cayleynets}, for example. In spite of this, spatial models are favored over spectral models for a variety of reasons, including efficiency, generality, and adaptability. To begin, the efficiency of spectral models is lower than that of spatial models. Either eigenvector computation or simultaneous handling of the entire graph is required by spectral models. Neither option is acceptable. Due to the fact that spatial models directly conduct convolutions in the graph domain through the use of information propagation, they are more scalable to huge graphs. Instead of performing the computation on the entire graph, it can be done in batches of nodes at a time. Second, spectral models that are based on a graph Fourier basis are not very good at generalizing to new graphs. They presume the graph to be static. Any change made to a graph's appearance would cause an alteration to the graph's eigenbasis. On the other hand, spatially based models carry out graph convolutions locally on each node, which allows weights to be easily shared across a variety of locations and structures. Third, spectral-based models can only be used to analyze undirected graphs because of this limitation.

\begin{figure}[H]
    \centering
    \includegraphics[width=\textwidth]{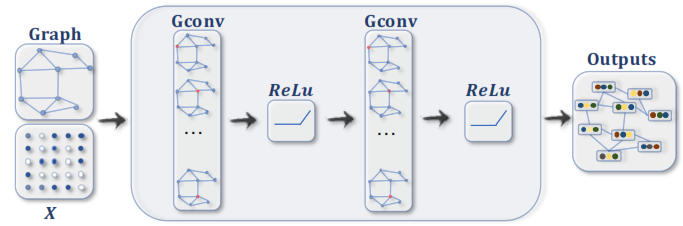}
    \caption{A ConvGNN for node classification.}\cite{wu2020comprehensive}
    \label{fig:ngcn}
\end{figure}

\begin{figure}[H]
    \centering
    \includegraphics[width=\textwidth]{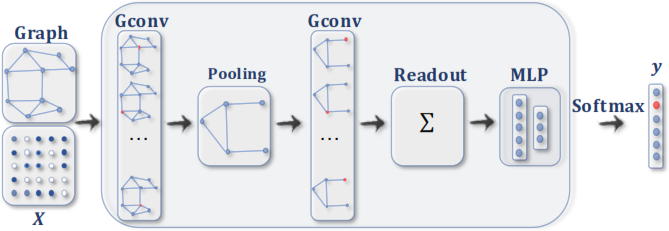}
    \caption{A ConvGNN for graph classification.}\cite{wu2020comprehensive,defferrard2016convolutional}
    \label{fig:ggcn}
\end{figure}

Because these graph inputs can be easily incorporated into the aggregation function, spatial-based models are more adaptable to handle multi-source graph inputs such as edge inputs \cite{scarselli2008graph,gilmer2017neural,kearnes2016molecular,pham2017column,simonovsky2017dynamic}, directed graphs \cite{atwood2016diffusion,li2017diffusion}, signed graphs \cite{derr2018signed}, and heterogeneous graphs \cite{such2017robust,wang2019heterogeneous}.
Figures \ref{fig:ngcn} and \ref{fig:ggcn} show different ConvGNNs used for node and graph classification respectively.

\subsubsection{Graph autoencoders}
\begin{figure}[!ht]
    \centering
    \includegraphics[width=\textwidth]{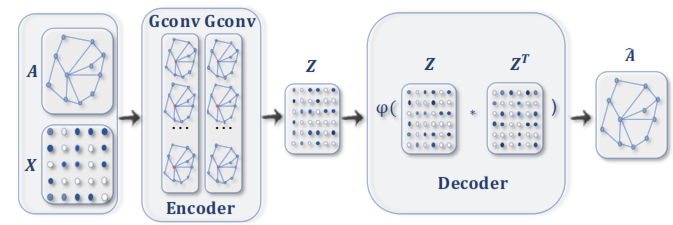}
    \caption{A GAE for network embedding.}\cite{wu2020comprehensive,kipf2016variational}
    \label{fig:gae}
\end{figure}

Graph autoencoders (GAEs) are frameworks for unsupervised learning that encode nodes and graphs into a latent vector space and then reconstruct graph data based on the encoded information. GAEs are put to use in the process of learning network embeddings as well as graph generating distributions. GAEs train latent node representations in order to accomplish network embedding by reconstructing graph structural information such as the graph adjacency matrix\cite{cao2016deep,kipf2016variational,wang2016structural,pan2018adversarially,tu2018deep,yu2018learning}. When it comes to graph generation, some methods build nodes and edges of a graph in a step-by-step fashion, while other methods immediately output a graph\cite{li2018learning,you2018graphrnn,simonovsky2018graphvae,ma2018constrained,de2018molgan,bojchevski2018netgan}. A Graph AutoEncoder for network embedding is shown in figure \ref{fig:gae}
\subsubsection{Spatial-temporal graph neural networks}
\begin{figure}[H]
    \centering
    \includegraphics[width=\textwidth]{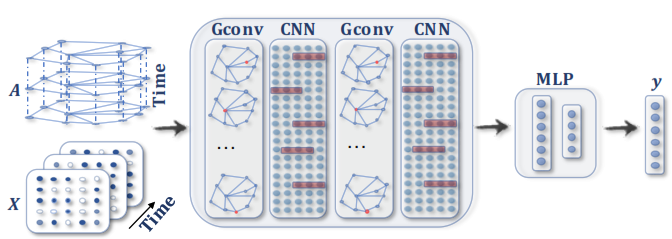}
    \caption{A STGNN  for graph forecasting.}\cite{wu2020comprehensive,yu2017spatio}
    \label{fig:STGNN}
\end{figure}

Spatial-temporal graph neural networks (STGNNs), are designed to discover hidden patterns within spatial-temporal graphs. These patterns are becoming increasingly important in a wide range of applications, including the forecasting of traffic speeds\cite{li2017diffusion}, the anticipation of driver maneuvers \cite{jain2016structural}, and the recognition of human actions \cite{yan2018spatial}. The aim behind STGNNs is to take into account both the geographical dependency and the temporal dependency of events simultaneously. Many of the systems being used now combine graph convolutions, which are used to describe spatial dependency, with either RNNs or CNNs, which are used to model temporal dependency\cite{seo2018structured,yu2017spatio,wu2019graph,guo2019attention}.
Figure \ref{fig:STGNN} shows an STGNN for spatial-temporal graph forecasting.

\chapter{Related Work}
\label{ch3}
In this chapter Graph neural networks different algorithms are introduced, then graph-based anomaly detection methods, then Heterogeneous aggregations for Graph Neural Networks, and finally, self-supervised graph learning which are all strongly related topics to this thesis.



\section{Graph Neural Networks}

GNNs have made significant strides forward in terms of their ability to learn graph representations.
The fundamental concept that underpins GNNs is that node representations can be kept up to date through the aggregation of messages from the various neighborhood groups. Models of GNNs that are considered to be state-of-the-art include GCN \cite{kipf2017semi}, GraphSAGE \cite{hamilton2017inductive}, GAT \cite{velickovic2018graph}, GIN \cite{xu2018powerful}, and many others. The knowledge on the neighborhood is accumulated in these models in different ways, which is another manner in which these models differ from one another.

\subsection{Graph Convolutional Networks}
\begin{figure}[!ht]
    \centering
    \includegraphics[width=\textwidth]{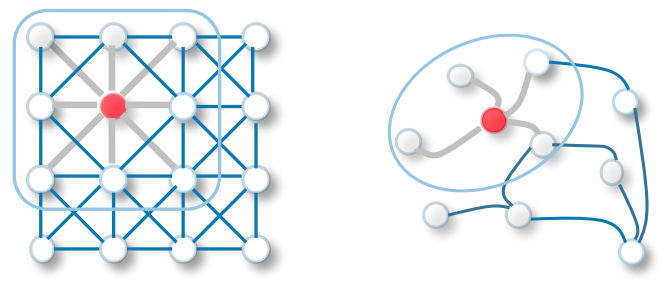}
    \caption{2D Convolution vs. Graph Convolution.} \cite{wu2020comprehensive}
    \label{fig:gcn}
\end{figure}

For instance, Graph Convolutional Networks (GCNs) \cite{kipf2017semi} uses a transductive approach to propagate messages that are derived from the graph Laplacian matrix.
GCNs are considered spectral ConvGNNs\cite{liu2020introduction}.
Convolution in GCNs is very similar to the convolution in neural networks previously discussed in \nameref{cnn}.
Similar convolution actions are carried out by GCNs, however the model learns the features by looking at the nearby nodes. The primary distinction between CNNs and GNNs is that the former were developed specifically to function on regular (Euclidean) organized data, whilst the latter are generalized CNNs with varying numbers of connections and unordered nodes (irregular on non-Euclidean structured data) as shown in \autoref{fig:gcn} \cite{wu2020comprehensive} where each pixel in an image is treated as a node, similar to a node in a graph, and its neighbors are determined by the filter size. The weighted average of the pixel values for the red node and its neighbors are used in the 2D convolution. A node's neighbors are arranged and have a set size.\cite{wu2020comprehensive}

So the embeddings are calculated for node $v$ at layer $k$ as follows:

\begin{equation}
h_v^{(k)}=\operatorname{ReLU}\left(W \operatorname{MEAN}\left(\left\{h_u^{k-1}, \forall u \in \mathcal{N}(v) \cup\{v\}\right\}\right)\right)
\end{equation}\label{eqn:gcn}
Where $W$ is weight matrix to be multiplied by mean of the node embeddings of the previous layer.

\subsection{GraphSAGE}
GraphSAGE (SAmple and aggreGatE) \cite{hamilton2017inductive} presents a paradigm for inductive learning in which an aggregation function such as mean, max, or LSTM builds node embeddings by aggregating messages from a node's local neighborhood, GraphSAGE is considered spatial ConvGNNs \cite{liu2020introduction}. 
\begin{figure}[!ht]
    \centering
    \includegraphics[width=\textwidth]{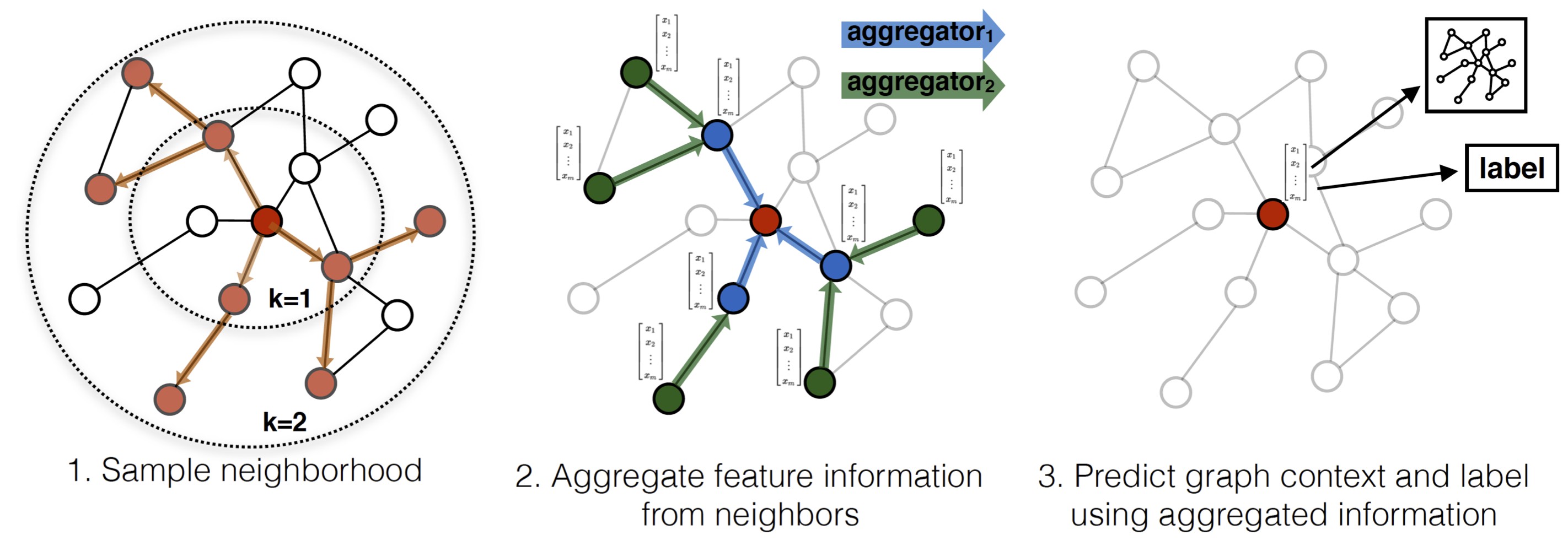}
    \caption{Diagrammatic representation of the GraphSAGE technique of sampling and aggregating graph data.} \cite{hamilton2017inductive}
    \label{fig:gsage}
\end{figure}
GraphSAGE operates under the presumption that nodes that are located in the same sample neighborhood should have embeddings that are similar\cite{hamilton2017inductive}, as shown in figure \ref{fig:gsage}; first step would be determining the neighborhood and more precisely, determining the extent of the neighborhood, a parameter known as k within the algorithm is responsible for determining the extent of the neighborhood. If k is equal to one, then only the nodes that are close to one another are considered to be comparable. In the event that K equals 2, the nodes located at a distance of 2 are also considered to be in the same neighborhood, this means a very large k will result in a similar embedding of all nodes; second step is the aggregation taking place after defining the neighbors as discussed in the first step, aggregators generate neighborhood embeddings based on the embeddings of the nodes and the weights which are hyperparameters or learnt; an advantage of inductive learning algorithm over the transductive learning algorithm deepwalk\cite{perozzi2014deepwalk} arises here as we have seen the node embeddings depend on neighbors or k value and node features meaning that when we have a new node no need for retraining the whole network to get an unseen or new node embedding, and as per the equations below the embeddings of node $v$ at layer $k$ are calculated.
\begin{equation}
\begin{aligned}
&a_v^{(k)}=\operatorname{MAX}\left(\left\{\operatorname{ReLU}\left(W h_u^{(k-1)}\right), \forall u \in \mathcal{N}(v)\right\}\right) \\
\end{aligned}
\end{equation}
\begin{equation}
\begin{aligned}
h_v^{(k)}=&W\left[h_v^{(k-1)}, a_v^{(k)}\right] 
\end{aligned}
\end{equation}
Where $W$ is weight matrix, h is embeddings vector.

\subsection{Graph Isomorphism Network}
Graph Isomorphism Network (GIN) \cite{xu2018powerful} uses a sum-like aggregation function, which has been shown to be just as effective as the Weisfeiler-Lehman graph isomorphism test \cite{weisfeiler1968reduction},this test is about deciding if 2 graphs are identical from a topological point of view (i.e. isomorphic), this can be determined after aggregating the labels of the nodes and their neighbors then transforms the aggregated labels into a set of distinct new labels via hashing and decide that the graphs are not isomorphic if the nodes labels are not the same between the graphs for the same iteration, back to the GIN algorithm which computes the embeddings of node $v$ at layer $k$ as per the equations below.
\begin{equation}
\begin{aligned}
&h_v^{(k)}=\operatorname{MLP}^{(k)}\left(\left(1+\epsilon^{(k)}\right) \cdot h_v^{(k-1)}+\sum_{u \in \mathcal{N}(v)} h_u^{(k-1)}\right) \\
\end{aligned}\label{eqn:gin}
\end{equation}
\begin{equation}
\begin{aligned}
&h_G=\operatorname{CONCAT}\left(\operatorname{READOUT}\left(\left\{h_v^{(k)} \mid v \in G\right\}\right) \mid k=0,1, \ldots, K\right)
\end{aligned}
\end{equation}
Where $\epsilon<<1$ or a learnable parameter, where node embeddings are initialized with features $h_v^{(0)}=x_v$, MLP is a multi-layer perceptron to add non-linearity, which makes the GIN act as a generalization to GraphSAGE and GCN.

\subsection{Graph Attention networks}
With the groundbreaking study Graph ATtention networks (GATs) \cite{velickovic2018graph}, graph attentive networks are researched to use attention mechanisms to give certain neighbors distinct weights using a self-attention technique.
Attention mechanisms introduce the attention coefficient that gives different neighbor nodes different weights, this idea is very important due to the reality that when we are updating the node embeddings we need to know which neighbor nodes should affect the embeddings of another node.
Attention coefficient $e_{i j}$ between nodes $i$ and $j$ with embeddings $h_i$ and $h_j$ respectively, can be defined as follows:
\begin{equation}
e_{i j}=a\left(W h_i, W h_j\right)
\end{equation}
but this formula results in non comparable weight so authors of GAT \cite{velickovic2018graph} used the softmax function to normalize the coefficients over all possible selections of $j$ in order to make the coefficients easily comparable across various nodes.
\begin{equation}
\alpha_{i j}=\operatorname{softmax}_j\left(e_{i j}\right)=\frac{\exp \left(e_{i j}\right)}{\sum_{k \in \mathcal{N}_i} \exp \left(e_{i k}\right)}
\end{equation}

In the studies conducted by Veličković and colleagues \cite{velickovic2018graph}, the attention mechanism $a$ is a single-layer feedforward neural network that is parameterized by a weight vector \overrightarrow{\mathbf{a}} and used the LeakyReLU nonlinearity (with negative input slope = 0.2). When everything is taken into account, the coefficients that the attention mechanism computes look like this:
\begin{equation}
\alpha_{i j}=\frac{\exp \left(\operatorname{LeakyReLU}\left(\overrightarrow{\mathbf{a}}^T\left[\mathbf{W} \vec{h}_i \| \mathbf{W} \vec{h}_j\right]\right)\right)}{\sum_{k \in \mathcal{N}_i} \exp \left(\operatorname{LeakyReLU}\left(\overrightarrow{\mathbf{a}}^T\left[\mathbf{W} \vec{h}_i \| \mathbf{W} \vec{h}_k\right]\right)\right)}
\end{equation}
Where T for transposed, and $||$ for concatenation.

These different graph neural network models or methods have seen widespread implementation in a variety of real-world applications, including anomaly detection and bioinformatics \cite{shi2020graph,shi2020graphaf} in addition to recommender systems \cite{he2020lightgcn,ying2018graph,wang2019neural}, amongst others.

\subsection{Graph Neural Network Heterogeneous Aggregations}
Leng et al. in their work \cite{leng2021enhance} showed how Heterogeneous Aggregations (HAG-Net)
provide several aggregating processes in each layer so that each layer can separately learn information from its neighboring layers resulting in better representation\cite{wei2022designing}.

For a graph $G=(V, E)$ with nodes $v, u \in V$ and edge $e_{u v} \in E$, The formulation of a generic layer with heterogeneous aggregations looks like this:
\begin{equation}
h_v=\psi\left(\mathbf{C}\left(h_v, \bigoplus_i^{M-1} \phi_i\left(\mathbf{A}_{i, u \in \mathcal{N}(v)}\left(\left\{\left(h_v, h_u, e_{u v}\right)\right\}\right)\right)\right)\right)
\end{equation}

, where $\mathbf{A}_i(\cdot), i=0 \cdots M-1$ are $M$ different operators for aggregation, $\bigoplus$ is the  operator to merge aggregation results for $M$ neighborhood, and center node $v$ 's feature are updated with the merged aggregation result by $\mathbf{C}(\cdot)$. $\psi$ and $\phi_i$ are non-linear or linear transform functions, and the layer index $k$ has been left out of this description for the sake of brevity. The left-hand calculation is always related to layer $k-1$, unless it is explicitly stated otherwise.

In most cases, the node representation $h_v$ from the last aggregate layer is the one that is used for prediction when dealing with node-level activities. READOUT is a function that gathers node features from the last aggregation layer to produce the complete graph's representation as a whole. This function is used for graph-level operations, in this case, the READOUT function also benefits from the improved information propagation brought about by the utilization of heterogeneous aggregations. Note that when using the READOUT function, aggregations are carried out across the entirety of the graph rather than just the immediate neighborhood.

\begin{equation}
h_G=\psi\left(\bigoplus_i^{M-1} \phi_i\left(\mathbf{A}_{i, v \in V}\left(\left\{h_v\right\}\right)\right)\right)
\end{equation}

\pagebreak
\section{Self-supervised Graph Learning}
In the field of computer vision, self-supervised learning, also known as pre-training, is a strategy that is widely used and proven to be effective\cite{chen2020simple,kolesnikov2019revisiting,zhang2016understanding} . Contrastive learning (CL), one of the SSL schemes, has recently attracted a lot of attention due to the surge of interest in visual representation learning\cite{chen2020simple,he2020momentum}. In a related vein, CL-based SSL schemes have also been looked into for use with graph data. The early attempts of unsupervised graph learning, such as GAE\cite{kipf2016variational}, GraphSage \cite{hamilton2017inductive}, node2vec \cite{grover2016node2vec}, deepwalk \cite{perozzi2014deepwalk} and LINE \cite{tang2015line}, which try to reconstruct the adjacency information of nodes, can be viewed as a kind of ``local contrast'' between a node and its neighbors in order to preserve the local homophily. This ``local contrast'' between a node and its neighbors is necessary in order to reconstruct the adjacency information of the later proposed GCC \cite{qiu2020gcc}, which designs subgraph-level instance discrimination to capture transferable local structural patterns, can be viewed as a ``local contrast'' between multiple views of nodes. This was done in order to capture transferable local structural patterns (sampled ego-networks). Motivated by DIM \cite{hjelm2018learning}, DGI \cite{velickovic2018deep} and InfoGraph \cite{sun2019infograph} have proposed contrasting the node graph with the global graph, which can be viewed as a ``local-global contrast'' to capture the information about the global structure. GPT-GNN \cite{hu2020gpt} is able to learn graph structures as well as the attributes of nodes through the use of a generative SSL scheme. Both You et al. and Hassani et al. \cite{you2020graph,hassani2020contrastive} investigated several distinct kinds of graph data augmentations that are applied to DGI-based SSL scheme.
The work done by Wang et al. \cite{wang2021decoupling} was distinct from the methods described above in two key respects. First, they investigated the efficacy of decoupled training on anomaly detection and found that inconsistency is a key factor that affects the quality of representation learning. This finding was a revelation for them. Despite the fact that Kang et al. \cite{kang2019decoupling} have also investigated the effectiveness of decoupled training in visual representation learning, their focus is on finding a solution to the problem of class imbalance. Wang et al. \cite{wang2021decoupling} were the first researchers to make the connection between the issue of inconsistent training and the use of decoupled training. Second, in order to compare and contrast the local and the semi-global representations, they propose using DCI.
\section{Modes of Anomaly Detection}
The anomaly detection has three modes of learning, either supervised, unsupervised or semi supervised.
\subsection{Supervised Anomaly Detection}
Anomaly Detection Through Supervision Techniques that are trained in supervised mode presuppose the availability of a training data set that contains cases that have been labeled for both the normal and the anomalous classes. Building a prediction model that differentiates between normal and abnormal classes is a common method taken in situations like this. Any previously unseen data instance is put through a comparison with the model to see which class it falls into. When performing supervised anomaly detection, you may run into two significant challenges. First, when compared to the normal occurrences in the training data, the number of anomalous examples is significantly lower. In the literature on data mining and machine learning, problems caused by imbalanced class distributions have been discussed \cite{joshi2001mining,joshi2002predicting,chawla2004special,phua2004minority,weiss1998learning,noble2003graph,vilalta2002predicting}. Second, it is typically difficult to acquire labels that are both accurate and representative, particularly for the anomaly class.\cite{theiler2003resampling,abe2006outlier,steinwart2005classification}. There have been a variety of methods that have been developed that inject false anomalies into a regular data set in order to acquire a labeled training data set. The supervised anomaly detection challenge is quite similar to the problem of creating predictive models, with the exception of these two difficulties.\cite{chandola2009anomaly}

\subsection{Unsupervised Anomaly Detection}
Techniques that function in an unsupervised mode don't need any sort of training data, therefore they can be applied to the widest variety of situations. The procedures that fall into this category operate under the tacit premise that regular occurrences in the test data are significantly more often than outliers. If it turns out that this assumption is incorrect, then the accuracy of such procedures is severely compromised.
By selecting a subset of the unlabeled data set to act as training examples, one can convert a number of semisupervised methods into ones that can function in an unsupervised manner. This adaptation works under the assumption that the test data contains relatively few outliers and that the model that was developed during training is resilient to the small number of outliers that are present.\cite{chandola2009anomaly}

\subsection{Semi-supervised Anomaly Detection}
Methods that function in a semisupervised mode begin with the presumption that the training data contains labeled instances exclusively for the normal class.
Unsupervised approaches provide a wider range of applications because they do not require any labels to be associated with the anomalous classes. For instance, in the research paper "Spacecraft Fault Detection" by Fujimaki et al. (2005)\cite{fujimaki2005approach}, an anomalous scenario might indicate an accident, which is difficult to model. Building a model for the class that corresponds to normal behavior and utilizing that model to find anomalies in test data is the common strategy that is employed in techniques of this kind.\cite{dasgupta2002anomaly,dasgupta2000comparison,warrender1999detecting} 
There is a restricted collection of anomaly detection approaches that operate under the assumption that there is availability of only the anomalous instances for training. These methods are not frequently utilized, partly because it is challenging to get a training data set that accounts for every conceivable anomalous behavior that can occur in the data. This is the primary reason why such techniques are not commonly employed.\cite{chandola2009anomaly}
\section{Graph-based Anomaly Detection}
Earlier studies found that dense block identification \cite{hooi2016fraudar,shah2014spotting,shin2016m}, iterative learning \cite{kumar2018rev2,li2012robust,mishra2011finding,wang2011review,wang2012identify,zhang2021tadoc}, and belief propagation on graphs \cite{akoglu2013opinion,rayana2015collective} were the most effective methods for spotting anomalies. However, these early attempts typically rely on the rules or features that were defined by humans, which makes it difficult to generalize the results to a number of different datasets. GNNs have been so successful in their applications that most modern algorithms automatically use them to summarize any unusual patterns they encounter. Examples include GAS \cite{li2019spam}, FdGars \cite{wang2019fdgars}, GraphConsis \cite{liu2020alleviating}, and CARE-GNN \cite{dou2020enhancing} for detecting review fraud, GeniePath \cite{liu2019geniepath} and SemiGNN \cite{wang2019semi} for detecting financial fraud, FANG \cite{nguyen2020fang} for detecting fake news, ASA \cite{wen2020asa} for detecting mobile fraud, and MTAD-GAT \cite{zhao2020multivariate} for detecting time-series anomalies. 

In order to address the issue of anomaly detection, these models modify the original GCN \cite{kipf2017semi}, the graph attention network \cite{velickovic2018graph,wang2019heterogeneous}, or the heterogeneous GNNs \cite{yang2020heterogeneous}. 
Current efforts have primarily concentrated on putting forth a powerful GNN encoder for node representation learning that is directed by the labels.
In these types of models, the processes of learning representations and classifying data are typically carried out simultaneously. 
In order to capture the desired patterns for anomaly detection, an appropriate graph SSL method is proposed in this study as another option for separating these two components based on the work of \cite{wang2021decoupling}.

\chapter{Methodology}
\label{ch4}
\theoremstyle{definition}
\newtheorem{definition}{Definition}
In order to gain an understanding of the suggested model, we will begin this chapter by providing a summary of the preliminary definitions. In addition, we will go over the baseline model in great depth, as well as the method of "Decoupling Representation Learning and Classification for GNN-based Anomaly Detection", and ultimately, we will go over the improved recommended design.

\section{Definitions and Notations}
\begin{definition} Graph: a graph or network $\mathcal{G}=(\mathcal{V}, \mathcal{E})$ is a mathematical representation defined by a set of nodes or vertices $\mathcal{V}$ and a set of edges or links $\mathcal{E}$ between these nodes, an edge $\mathcal{E}$ directed from node $ u \in \mathcal{V}$ to node $ v \in \mathcal{V}$ is denoted as $ (u,v) \in \mathcal{V}$ .\cite{hamilton2020graph,anzai2012pattern}
\end{definition}

\begin{definition}Simple graph: a graph or network where there is a maximum of one edge connecting each pair of nodes, where there are no edges connecting a node to itself, and where all of the edges are undirected, i.e., $(u, v) \in \mathcal{E} \leftrightarrow(v, u) \in \mathcal{E}$.\cite{hamilton2020graph,cai2018comprehensive}
\end{definition}

\begin{definition}Adjacency matrix : An adjacency matrix $\mathbf{A} \in$ $\mathbb{R}^{|\mathcal{V}| \times|\mathcal{V}|}$ provides a straightforward and practical method for representing graphs. In order to use an adjacency matrix to represent a graph, the nodes in the graph must first be sorted in such a way that each node references a specific row and column in the adjacency matrix. The existence of edges can then be represented as entries in this matrix $\mathbf{A}[u, v]=1$ if $(u, v) \in \mathcal{E}$ and $\mathbf{A}[u, v]=0$ otherwise.\cite{hamilton2020graph}
\end{definition}

\begin{definition}Directed graphs: they are types of graphs where edge direction is important and consequently, adjacency matrix $\mathbf{A}$ does not automatically have to be symmetric, for the undirected graphs the the edges are bidirectional meaning they have no direction.
\cite{hamilton2020graph,goyal2018graph}
\end{definition}

\begin{definition}Weighted graph: a weighted graph will contain weighted edges representing the strength between connected nodes, which means that the elements in the adjacency matrix will be arbitrary real values rather than the binary values 0 and 1, on the other hand weighted graphs are called binary graphs as they have either 0 or 1 values for edges in adjacency matrix $\mathbf{A}$. \cite{hamilton2020graph,xu2020identifying}
\end{definition}

\begin{definition} Heterogeneous graph: is a graph that has more than one type of nodes or more than one type of edges.\cite{cai2018comprehensive}
\end{definition}
\begin{definition} Homogeneous graph: is a graph that its nodes are all of one type of nodes and its edges are all of one type of edges. \cite{cai2018comprehensive}
\end{definition}
\begin{definition}
First-order proximity: weight of an edge between nodes can be considered the first-order proximity.\cite{cai2018comprehensive}
\end{definition}
\begin{definition}
second-order proximity: similarity between a node's neighborhood to another node's neighborhood is the second order proximity.\cite{cai2018comprehensive}
\end{definition}
\begin{definition}Joint training: nodes are estimated by $\hat{Y}=f(g(G, \theta), \phi)$ and $\theta and \phi$ are trained using one supervised loss function $\mathcal{L}_{SL}$ \cite{wang2021decoupling}.
\end{definition}
\begin{definition}Decoupled  training: In decoupled learning the representation learning and the classification are decoupled. which means the learning process is done on 2 steps, in the first step the node embeddings are estimated by $H = g(G,\theta)$ and $\theta$ is trained by a self-supervised loss function $\mathcal{L}_{SSL}$ as it is not depending on node labels that are observed $Y^{L}$. In the second step, the node labels are estimated by $\hat{Y}=f(g(G, \theta), \phi)$, and trains $\phi$ using a supervised loss function $\mathcal{L}_{SL}$ based on the $\theta$ learned in the first step, in the second step the $\theta$ gets tuned also\cite{wang2021decoupling}.
\end{definition}
\begin{definition} Anomaly detection: it is a practice of identifying data instances that considerably diverge from the majority of other data instances. \cite{chalapathy2019deep,pang2021deep}
\end{definition}

\section{Baseline approach}
In this section the different approaches developed by Wang et al. \cite{wang2021decoupling} whether joint or decoupled training for GNN-based anomaly detection.
\subsection{Joint learning}
\begin{figure}[!ht]
    \centering
    \includegraphics[width=\textwidth]{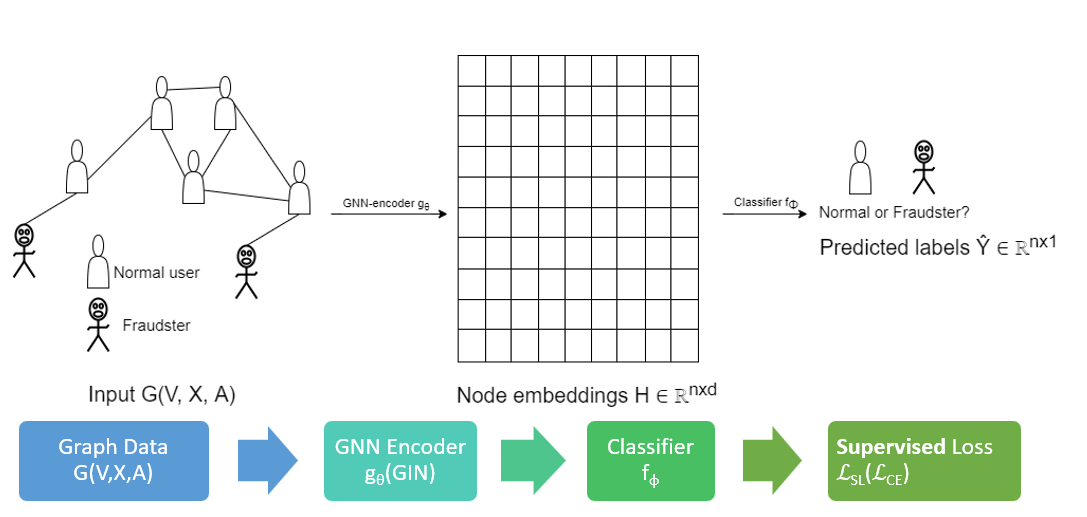}
    \caption{Joint training block diagram}
    \label{fig:joint}
\end{figure}

In this part an introduction to the four forming parts of the joint training namely, Graph Data Input, GNN encoder, classifier and at the end the supervised loss components are introduced, a representation of different components is shown in figure \ref{fig:joint}.
\subsubsection{Input Data}
The data graph we have as input are partially labeled nodes $Y^L$ where we have labels $y\in \{ 0,1 \}$, to label fraudulent with 1 or not fraudulent with 0. and the adjacency matrix $A$, note that our graphs are Heterogeneous graphs from a side we have users and on the other side different objects depending on each dataset for instance, subreddit in case of reddit dataset and products for amazon dataset etc., using the file representing the adjacency matrix, the features matrix $X \in R^{n \times d}$ where $n$ is number of nodes and d is 64.
\subsubsection{GNN Encoder}
GIN was chosen as a GNN encoder, recall the GIN node embeddings equation \ref{eqn:gin} \begin{equation}
\begin{aligned}
&h_v^{(k)}=\operatorname{MLP}^{(k)}\left(\left(1+\epsilon^{(k)}\right) \cdot h_v^{(k-1)}+\sum_{u \in \mathcal{N}(v)} h_u^{(k-1)}\right) \\
\end{aligned}
\end{equation}
\begin{equation}
\begin{aligned}
&h_G=\operatorname{CONCAT}\left(\operatorname{READOUT}\left(\left\{h_v^{(k)} \mid v \in G\right\}\right) \mid k=0,1, \ldots, K\right)
\end{aligned}
\end{equation}
Where $\epsilon<<1$ or a learnable parameter, where node embeddings are initialized with features $h_v^{(0)}=x_v$, MLP is a multi-layer perceptron to add non-linearity.
\subsubsection{Classifier}\label{baseline:classifier}
The classifier used is simply based on linear mapping of node embeddings then a non linear activation function is applied to this linear mapping, as shown in the equation below.
\begin{equation}
p_i=\sigma\left(W^{\top} \boldsymbol{h}_i^{(K)}+b\right)
\end{equation}
where $p_i$ anomaly score for node $i$, $W$ and $b$ are learnable parameters, ${h}_i^{(K)}$ is final node representation, $\sigma$ is the sigmoid activation function recall equation \ref{eqn:simoid}, $p_i$ is the abnormal score of node $v_i$.
\subsubsection{Supervised Loss} \label{baseline:SLCE}
The authors Wang et al. \cite{wang2021decoupling} chose cross-entropy \begin{equation}
\mathcal{L}_{SL}=\mathcal{L}_{C E}=-\frac{1}{\left|Y^L\right|} \sum_{i=1}^{\left|Y^L\right|}\left(y_i \cdot \log p_i+\left(1-y_i\right) \cdot \log \left(1-p_i\right)\right)
\end{equation}
Where $Y^L$ number of labeled nodes, $y_i$ is the label, $p_i$ is the node anomaly prediction.
This cross-entropy supervised loss function is used to train the aforementioned classifier.
\subsection{Decoupled learning}
\begin{figure}[!ht]
    \centering
    \includegraphics[width=\textwidth]{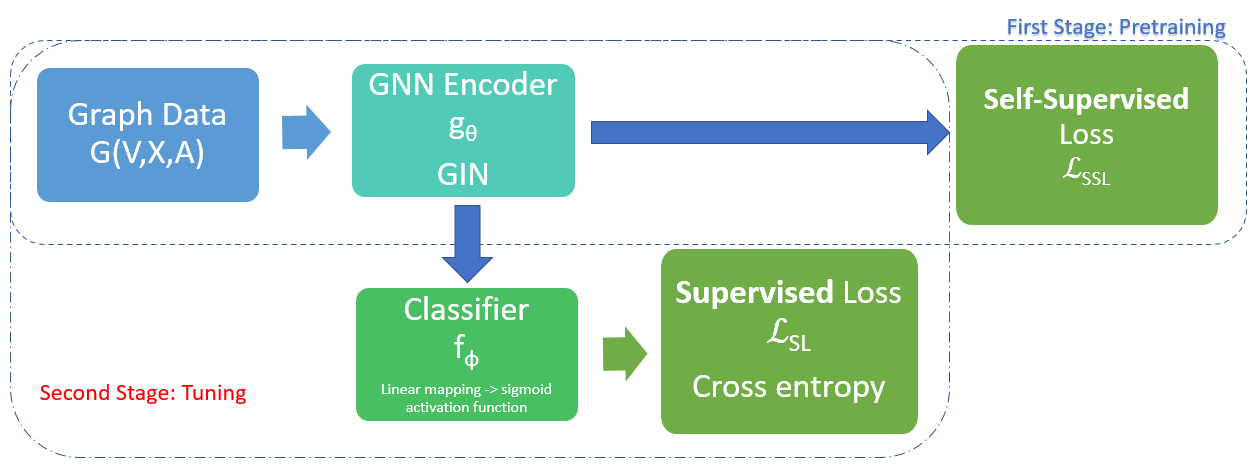}
    \caption{Decoupled training block diagram}
    \label{fig:decoupled}
\end{figure}

For decoupled learning all the blocks used for joint learning are utilized for decoupled learning in addition to the self-supervised loss. \cite{velickovic2018deep} as shown in figure \ref{fig:decoupled}.
\subsubsection{Self-supervised loss}\label{baseline:ssl}
\begin{figure}[!ht]
    \centering
    \includegraphics[width=0.8\textwidth]{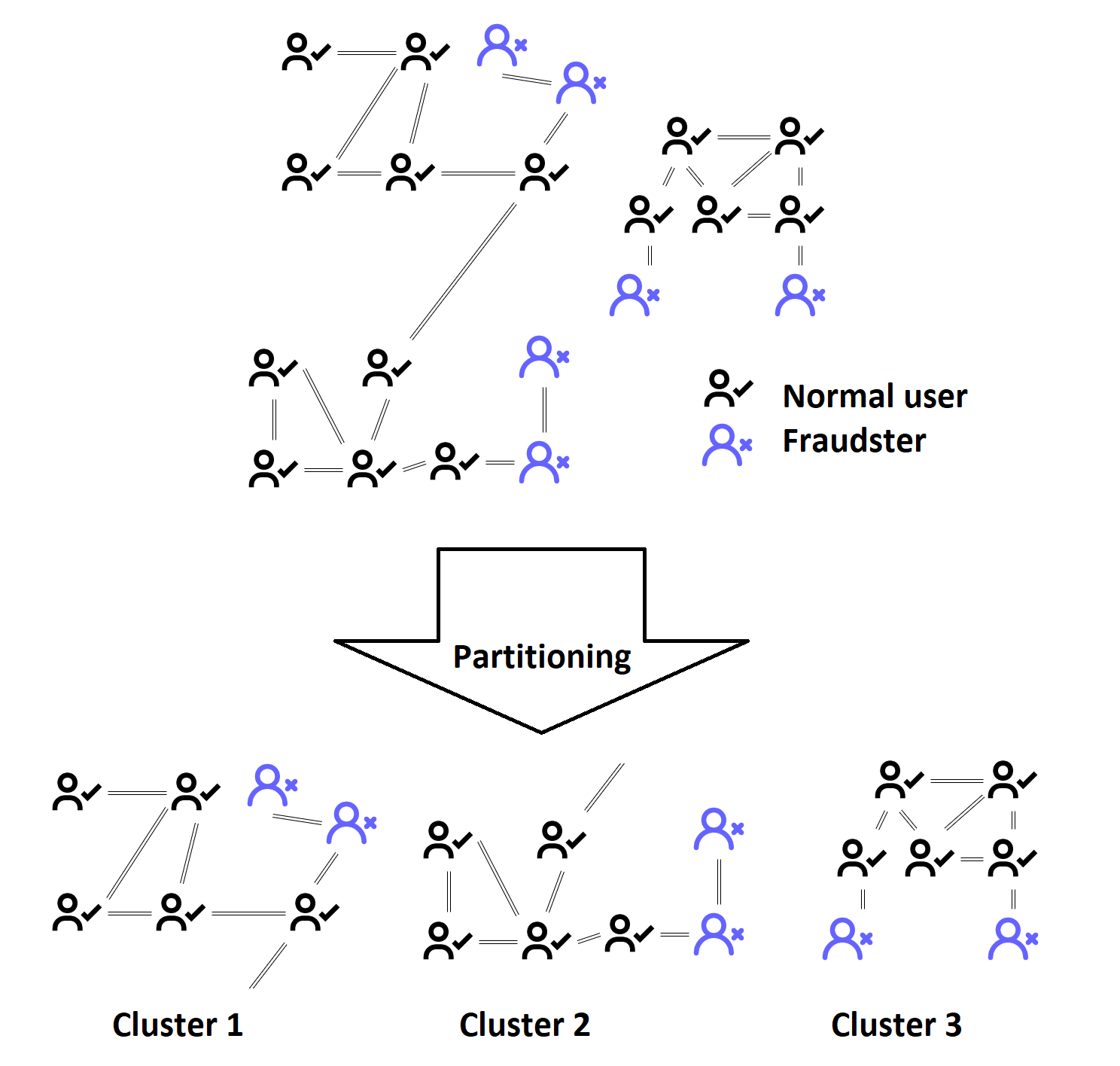}
    \caption{Deep Cluster Infomax Graph clustering}
    \label{fig:clust}
\end{figure}

The self-supervised loss chosen by Wang et al. \cite{wang2021decoupling} to be the deep graph infomax based on the anomaly definition and then they proposed a better one which is deep cluster infomax.
\paragraph{Deep Graph Infomax (DGI)}: This module makes use of the graph structure to understand the patterns in the data that are associated with each node. More specifically, it encodes the global information into node representations in order to represent both the individual behavior patterns and the normal pattern that is occupied by the majority of the nodes.
\begin{equation}
\mathcal{L}_{S S L}=\mathcal{L}_{D G I}=-\frac{1}{2 n} \sum_{i=1}^n\left(\mathbb{E}_G \log \mathcal{D}\left(\boldsymbol{h}_i^{(L)}, s\right)+\mathbb{E}_{\tilde{G}} \log \left(1-\mathcal{D}\left(\tilde{\boldsymbol{h}}_i^{(L)}, s\right)\right)\right)
\end{equation}
Where $\mathbb{E}$ is expectation over a graph, $G$ is the positive samples graph, $\tilde{G}$ is the negative samples graph, $h^(L)$ node embedding, $\mathcal{D}$ is discriminator, in a case that each of the positive samples graph and negative sample graphs has $n$ nodes, and $s$ is graph summary vector is calculated according to the following equation.
\begin{equation}
s=\sigma\left(\frac{1}{n} \sum_{i=1}^n h_i^{(L)}\right)
\end{equation}
\paragraph{Deep Cluster Infomax (DCI)}\label{baseline:dci}: The Graph G is partitioned into K clusters using K-means algorithm according to node features X as shown in figure \ref{fig:clust}, after each re-clustering training epoch, the nodes undergo a process that re-clusters them based on the embeddings of the nodes.
Following the clustering process, the cluster-level representation $S_k$ is calculated for different clusters in order to summarize the behavior of the majority within cluster k: $C_k$ contains $n_k$ nodes.

\begin{equation}
s_k=\sigma\left(\frac{1}{n_k} \sum_{v_i \in V_k} \boldsymbol{h}_i\right)
\end{equation}
\begin{equation}
\mathcal{L}_{D C I}^k=-\frac{1}{2 n_k} \sum_{v_i \in V_k}\left(\mathbb{E}_{C_k} \log \mathcal{D}\left(\boldsymbol{h}_i, s_k\right)+\mathbb{E}_{\tilde{C_k}} \log \left(1-\mathcal{D}\left(\tilde{\boldsymbol{h}}_i, s_k\right)\right)\right)
\end{equation}
Where $\mathbb{E}$ is expectation over a cluster k, $C_k$ is the positive samples cluster, $\tilde{C}$ is the negative samples graph, $h^(L)$ node embedding, $\mathcal{D}$ is discriminator, in a case that each of the positive samples graph and negative sample graphs has $n$ nodes, and $s$ is graph summary vector is calculated according to the following equation. 
\begin{equation}
\mathcal{L}_{D C I}=\frac{1}{K} \sum_{k=1}^K \mathcal{L}_{D C I}^k
\end{equation}
\pagebreak
\section{Proposed approach}

\begin{figure}[!ht]
    \centering
    \includegraphics[width=\textwidth]{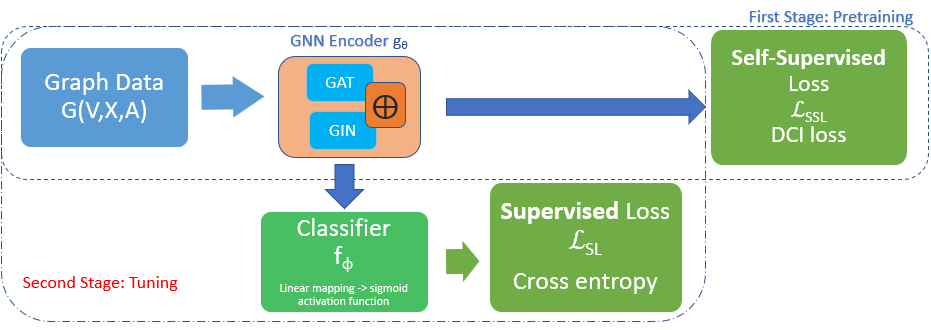}
    \caption{Proposed design changing the GNN encoder}
    \label{fig:dsgn}
\end{figure}

This thesis proposed approach that is pretty similar to where most of the researchers think of which is building a more powerful GNN encoder as GIN is the most powerful GNN encoder this was not an easy direct task to change the GNN encoder, and influenced by the idea that a weak learner that is paired with other weak learners to generate a strong learner, instead of having one pure more powerful encoder we tried to have multiple encoders, and averaging the embeddings got from different encoders for a better representation learning and normalizing or standardizing the node features.
GIN\cite{xu2018powerful} and GAT\cite{velickovic2018graph} encoders and their results are chosen to be aggregated with an average aggregation function, when this approach is experimented it shows results better than using pure GIN encoder solely as shown in figure \ref{fig:dsgn}.
GCN\cite{kipf2017semi} is known of aggregating neighbor features in addition to self features, the attention in case of GCN is depending only on graph structure and not the node features,  on the other hand hidden features are subjected to a local averaging operation by GAT. However, rather than relying solely on the graph structure to direct the propagation as in the case of GCN, it uses a learnable function called alpha to reweight the propagation weight such that it takes into account the hidden features.
GIN \cite{xu2018powerful} also has great empirical representational power due to the fact that it fits the training data almost flawlessly, in contrast to the less powerful GNN variations, which frequently drastically underfit the training data.
So combining the GIN and GAT with an average aggregation function results in capturing the full extent of node information and achieving enhanced node representation as it efficiently represent and preserve the information about the nodes while also maintaining the features of the network.\\

As per the proposed design shown in figure \ref{fig:dsgn}, the other building blocks of the baseline were not changed so namely, the self-supervised loss was of the \nameref{baseline:dci}, the classifier is still very same to the aforementioned \nameref{baseline:classifier} in the baseline approach that learns with the same cross-entropy \nameref{baseline:SLCE}.
The code for the proposed implementation was uploaded to GitHub \cite{Hafez_Global_Context_Enhanced}.
\subsection{Other experimented approaches}
In this section I will give a short summary of other trials of modifying the proposed approach, but there results turned to be sub-optimum.

\subsubsection{Joint training with GAT}
As expected because of what is reported by \cite{wang2021decoupling}, testing the joint training with GAT encoder was giving sub-optimal results or lower average AUC for most of the datasets, this shows that for joint training the GIN was the best GNN encoder.

\subsubsection{Modifying the DCI clustering algorithm}
The self supervised loss for the GNN encoder used was the aforementioned DCI loss which uses k-means, as a modification affinity propagation \cite{frey2007clustering} was considered, but the result of this modification were disappointing it failed due to the fact that the complexity of Affinity Propagation is considered to be its primary shortcoming. The temporal complexity of the algorithm is on the order of $O(N^2T)$, where $N$ is the number of samples and $T$ is the number of iterations performed until convergence is achieved. In addition, the memory complexity is of the order $O(N^2)$ if a dense similarity matrix is employed, but it is reducible if a sparse similarity matrix is utilized. Because of this, Affinity Propagation works best with datasets that are either small or medium in size \cite{scikit-learn}, On the other hand, The k-means clustering algorithm has an average complexity that can be expressed as $O(knT)$, where $n$ is the total number of samples and $T$ is the total number of iterations\cite{scikit-learn}.

\subsubsection{Decoupled training training with different encoders}
Instead of the GAT and GIN encoders with average aggregation, different combinations of encoders were experimented namely, GCN with GIN and GAT with GCN but all these experiments lead to sub-optimal results i.e. less average AUC than the baseline approach, so these approaches were discarded.
\chapter{Experiments}
\label{ch5}
In this chapter, experiments to evaluate the viability of the suggested methodology are discussed. In order to accomplish this goal, Section \ref{sec:datasets} provides a comprehensive overview of the datasets that were utilized. A summary of the preprocessing methods that were necessary for the training of the models can be found under Section \ref{sec:prep}. A discussion of the experimental setups and environment that were utilized in the process of training and validating the models can be found in Section \ref{sec:imp}. For evaluation Area Under the Receiver Operating Characteristics measure was used and a short summary about the measure is presented in section \ref{sec:evaluation}. In the final section, Section \ref{sec:results}, the various experimental outcomes are broken down and discussed in detail.
\section{Datasets}\label{sec:datasets}
In the context of this thesis, a comparison of the baseline methodology and the proposed approaches is carried out using real-world user-object five established graph data sets. This comparison is carried out in order to demonstrate the viability of the proposed approaches. Table \ref{table:Datasets} contains a listing of the detailed statistics regarding these datasets. Each dataset that was used came from publicly accessible sources on the internet, and it is open for other people to make use of it. An overview of each of the datasets is provided in the following paragraphs.

\subsection{Reddit}
A user-subreddit graph on Reddit \cite{kumar2019predicting} is created by compiling all of the posts made by users on subreddits over the course of one month. This dataset contains ground-truth labels of users who have been banned from using Reddit. The Reddit dataset contains 10,000 (most active) users, 984 (most active subreddits or) objects, 10,984 nodes and 78,516 edges.

\subsection{Wiki}
The Wiki \cite{wang2021decoupling,snapnets,kumar2019predicting} public graph dataset contains the changes made to Wikipedia pages over the course of one month. It includes public ground-truth labels of users who have been banned. As items, the 1,000 (objects) most frequently edited pages were chosen, and editors who had made at least 5 edits were chosen as users (a total of 8,227 users), 9,227 nodes and 18,257 edges.

\subsection{Bitcoin Alpha}
Bitcoin Alpha \cite{wang2021decoupling,snapnets,kumar2018rev2,kumar2016edge} is a trust graph between users of the cryptocurrency Bitcoin who trade on the platform Alpha. The graph is made into a bipartite structure by dividing each user into a 'rater' with all of its outgoing edges and an 'object' with all of its incoming edges respectively. This dataset contains 214 users who have been assigned labels, 83 of them as anomaly i.e. 38.79\%, 3,286 is the total number of users, 3,754 objects, 7,040 nodes and 24186 edges.

\subsection{Amazon}
Amazon \cite{wang2021decoupling,snapnets,kumar2018rev2,mcauley2013amateurs} is a user-product rating graph dataset, and the edges of this graph describe the rating behaviors of users. Helpfulness votes are used to define ground truth, which is an indicator of malicious behavior. Users have more than 49 votes are considered non-malicious if the fraction of helpful-to-total votes is greater than 0.75, and fraudulent if the fraction is less than 0.25. This dataset contains 278 users who have been assigned labels, 23 as anomaly i.e. 8.27\%, total number of users (labeled and unlabeled) is 27,197 user, 5,830 objects, 33,027 nodes and 52,156 edges.

\subsection{Intrusion Detection Evaluation Dataset}
Intrusion Detection Evaluation Dataset (CIC-IDS2017)\cite{sharafaldin2018toward} it consists of different datasets for attacks took place at different workdays between Monday 03.07.2017 at 9:00 and Friday 07.07.2017 at 17:00 at morning and afternoon, this study considers only Thursday-Morning as other dataset were too big to process or had anomalies share that violate the anomaly definition.
The CIC-IDS2017 ``TrafficLabelling'' file had different datasets at different working hours namely, Monday, Tuesday, each in one csv file, then we have two csv files for Thursday as one represents Morning working hours and the second represent the afternoon working hours, for Friday we had three files one to represent the morning working hours and two to represent the afternoon working hours one of these two was representing the DDos and last file was representing the port scan attacks, last two files represent attacks took place on Friday 07.07.2017 afternoon.

\begin{table}[!ht]
\begin{tabular}{l|llll}
Dataset & \#Users(\% normal, abnormal) & \#Objects & \#Nodes & \#Edges \\ \hline
Reddit  & 10,000 (96.34\%, 3.66\%)     & 984       & 10,984  & 78,516  \\
Wiki    & 8,227 (97.36\%, 2.64\%)      & 1,000     & 9,227   & 18,257  \\
Alpha   & 3,286 (61.21\%, 38.79\%)     & 3,754     & 7,040   & 24,186  \\
Amazon  & 27,197 (91.73\%, 8.27\%)     & 5,830     & 33,027  & 52,156  \\
IDS2017 & 36,921 (94.58\%, 5.42\%)     & 17,787    & 54,708  & 79,462 
\end{tabular}
\caption{Datasets statistics}
\label{table:Datasets}
\end{table}

\section{Datasets Preprocessing}\label{sec:prep}
The first four datasets namely Reddit, Wiki, Alpha and Amazon were already preprocessed, each data set consists of two files: the first is the dataset\_label.txt which consists of a list of users starting from zero till number of users minus one and a label of this user 0 if it is normal and 1 in case of abnormal i.e. it is a matrix of the dimension number of users * 2 or $Y^L\in R^{U\times2}$, where U represents number of users; second file is showing the user-object relation i.e. adjacency matrix  its dimension is number of edges*2 or $A \in R^{m \times 2}$ where $m$ represents number of edges, user numbering or ID starts from zero (0) till number of users minus one and objects ID start from number of users as the first object id till number of objects in addition to number of users minus one which is the total number of nodes minus one.
\subsection{Preprocessing of IDS 2017 dataset}
For the IDS 2017 One csv file out of the files represent other days and working hours was selected for experiments namely the Thursday morning, if the label of an instance was ``Web Attack'' it was treated as fraud with ``1'' label, if it is labeled as ``BENIGN'' then it was labeled as not fraud with a ``0'' label
\section{Implementation}\label{sec:imp}
The proposed implementation is based on the baseline Deep Cluster infomax implementation for Decoupling Representation Learning and Classification for
Graph neural network based Anomaly Detection designed by Wang et al. \cite{wang2021decoupling}, adding to it the idea of multi-encoder aggregation influenced by Leng et al. work to Enhance Information Propagation for Graph Neural Network by Heterogeneous Aggregations \cite{leng2021enhance}.

\subsection{Parameter settings}
For all models, the input feature dimension is 64, the node representation dimension is 128, the number of GNN layers is 2, and the optimizer is Adam. The value of $\epsilon$ in for GIN equation \ref{eqn:gin} has been set to 0. The value of $\overline{t}$, which denotes the number of reclustering epochs, is going to be 20. The number of training epochs for the SSL pre-training has been fixed at 50 throughout the process.

Due to the fact that too many training epochs can be detrimental to the performance of this model, the early stopping strategy was used in the decoupled training with RW.

For the purpose of classification, the best testing result after one hundred iterations on each fold was recorded, and then the best summed AUC score across all of the folds was provided.
\subsection{Code implementations}
It was decided to employ open-source implementations for the GIN, and GAT GNN encoders, as well as for the DGI and DCI deep infomax\cite{velickovic2018graph,velickovic2018deep,wang2021decoupling,xu2018powerful,NEURIPS2019_9015}.
These codes have been altered so that they are more suited to the activities.
The code implementation of the proposed model is uploaded to GitHub\cite{Hafez_Global_Context_Enhanced}.

\section{Evaluation Metrics}\label{sec:evaluation}
In this section, an investigation is conducted into the usage of the area under the receiver operating characteristic (ROC) curve (AUC) as a performance measure for the learning algorithms\cite{bradley1997use}.
To understand the different parameters that build the ROC curve we start with a short definition of each value of the confusion matrix shown in Table \ref{table:confm}.
\begin{table}[!ht]
\centering
\resizebox{\columnwidth}{!}{%

\begin{tabular}{|l|l|l|l|} 
\cline{1-3}
\multirow{2}{*}{\begin{tabular}[c]{@{}l@{}}\textbf{True}\\\textbf{ Class}\end{tabular}} & \multicolumn{2}{c|}{\textbf{Predicted Class}}                                                                                                                                                               & \multicolumn{1}{l}{\textbf{}}                                     \\ 
\cline{2-3}
                                                                                        & negative                                                                                           & positive                                                                                               & \multicolumn{1}{l}{}                                              \\ 
\hline
negative                                                                                & \begin{tabular}[c]{@{}l@{}}TN: True negative\\correct rejection\end{tabular}                       & \begin{tabular}[c]{@{}l@{}}FP: False Positive\\type I error, false alarm,\\overestimation\end{tabular} & \begin{tabular}[c]{@{}l@{}}CN:\\Real negative cases\end{tabular}  \\ 
\hline
positive                                                                                & \begin{tabular}[c]{@{}l@{}}FN: False Negative\\ type II error, miss,\\underestimation\end{tabular} & \begin{tabular}[c]{@{}l@{}}TP: True Positive\\hit\end{tabular}                                         & \begin{tabular}[c]{@{}l@{}}CP:\\Real positive cases\end{tabular}  \\ 
\hline
\multicolumn{1}{l|}{}                                                                   & PN: Predicted negative                                                                            & PP: Predicted Positive                                                                                 & N: Total population                                               \\
\cline{2-4}
\end{tabular}}
\caption{Confusion Matrix}
\label{table:confm}
\end{table}

\subsection{True Positive Rate}
True Positive Rate (TPR), sensitivity, recall or hit rate
\[ TPR=\frac{TP}{CP} \]
\subsection{False Positive Rate}
\[ FPR=\frac{FP}{CN} \]
\\
ROC curve plots the sensitivity against FPR, and the AUC is the area under this curve that indicates the degree to which the model is able to differentiate between the two different classes i.e. The closer the AUC to one, the more accurately the classifier model can predict that 0 classes will be true 0 and 1 classes will be true 1, Ideally AUC is equal to 1 which will mean that the true positive and true negative distributions do not overlap, worst case scenario would not be when the AUC is 0 but when it is 50\%, when AUC is zero this means that the model is predicting the zeros as ones and vice versa, but when AUC is 50\% this means the classifier can't distinguish between zeros and ones.
\section{Results}\label{sec:results}
In this part a summary of the results obtained for this study with the design proposed above.
In an E-mail communication with the \cite{wang2021decoupling} author confirming that the results might change due to different versions of packages, I had to run the author's code on their datasets and my dataset to get the baseline results and did not depend on other reported results based on different packages and computing powers, so as to compare my design with the baseline design on the same environment.
The averaged AUC score is average AUC over 10-folds except for Amazon dataset it was 5-folds this due to that Amazon data set had few fraud labeled data points.
There were a difference between the published results of Wang et al. \cite{wang2021decoupling} and the results obtained after running the code, in an email communication with the author, they confirmed that it is possible for the k-fold split and the initialization of model parameters to be different for a variety of reasons, including but not limited to various random seeds, different versions of pytorch, different devices, and so on. Therefore, the outcomes of the experiment could be different in some way.

Averaged AUC results are shown in table \ref{table:aucresult}, and table \ref{table:time} show the time required for each model or algorithm.
\subsection{Source of gain}
The results in table \ref{table:aucresult}, suggest that the source of gain is coming from aggregating the results of GIN and GAT embeddings with an average function, when we used a solely encoder whether GIN or GAT we did not get the high values we received when we used GAT and GIN with an average aggregation function to generate the embeddings.

\begin{table}[!ht]
\resizebox{\columnwidth}{!}{%

\begin{tabular}{l|lllll}
                                                                                  & Reddit & Wiki   & Alpha  & Amazon & IDS2017 \\ \hline
\begin{tabular}[c]{@{}l@{}}Joint GIN\\ \end{tabular}     & $71.49\pm0.4$ & $72.60\pm0.3$ & $89.5\pm0.8$ & $73.3\pm0.7$ & $92.20\pm0.6$  \\
\begin{tabular}[c]{@{}l@{}}Joint GAT\\ \end{tabular}     & $73.49\pm 0.7$ & $70.44\pm 0.2$ & $86.48\pm 0.8$ & $70.27\pm 0.3$ & $87.45\pm 0.7$  \\
\begin{tabular}[c]{@{}l@{}}Decoupled DGI\\ \end{tabular} & $73.20\pm 0.5$ & $73.46\pm0.9$ & $89.40\pm0.7$ & $74.40\pm0.5$ & $92.87\pm0.8$  \\
\begin{tabular}[c]{@{}l@{}}Decoupled DCI (GIN)\\ \end{tabular} &
  {$73.18\pm 0.6$} &
  $\mathbf{74.81\pm 0.4}$ &
  $88.84\pm0.2$ &
  $75.92\pm0.3$ &
  $92.28\pm0.1$ \\
Decoupled DCI (GAT)                                                               & $67.17\pm0.8$ & $69.48\pm0.7$ & $89.20\pm0.5$ & $73.45\pm0.6$ & $87.24\pm0.4$  \\

\begin{tabular}[c]{@{}l@{}}MultiEncoder\\ (Our results)\end{tabular} &
  $\mathbf{74.44\pm 0.7}$ &
  {$72.99\pm 0.3$} &
  $\mathbf{89.56\pm 0.5}$ &
  $\mathbf{78.12\pm 0.3}$ &
  $\mathbf{92.91\pm0.4}$
\end{tabular}%
}
\caption{AUC Results as percentages}
\label{table:aucresult}
\cite{wang2021decoupling}
\end{table}
First four implementations are following Wang et al. baseline design\cite{wang2021decoupling}

According to the average AUC results and execution time results shown in tables \ref{table:aucresult} and \ref{table:time} respectively, it is clear that the proposed model is achieving better average AUC than any other model and converges far away faster than any other model.

\begin{table}[!ht]
\center
\resizebox{\columnwidth}{!}{%

\begin{tabular}{l|lllll}
                                                                                  & Reddit & Wiki  & Alpha & Amazon & IDS2017 \\ \hline
\begin{tabular}[c]{@{}l@{}}Joint GIN\\ \end{tabular}     & $3.31\pm 0.53$   & $2.54\pm 0.44$  & $1.92\pm 0.043$  & $2.34\pm 0.42$   & $14.63\pm 1.23$    \\
\begin{tabular}[c]{@{}l@{}}Decoupled DGI\\ \end{tabular} & $3.84\pm0.62$   & $2.91\pm0.75$  & $2.22\pm 0.41$  & $3.54\pm 0.73$   & $18.92\pm 1.73$    \\
\begin{tabular}[c]{@{}l@{}}Decoupled DCI\\ \end{tabular} & $3.84\pm 0.44$   & $2.97\pm 0.24$  & $2.25\pm 0.31$  & $5.73\pm 0.44$   & $17.35\pm 1.44$    \\
\begin{tabular}[c]{@{}l@{}}MultiEncoder\\ (Our results)\end{tabular}              & $\mathbf{0.77\pm 0.05}$  & $\mathbf{0.34\pm 0.01}$ & $\mathbf{0.36\pm 0.02}$ & $\mathbf{0.73\pm 0.05}$  & $\mathbf{1.93\pm 0.05}$
\end{tabular}}
\caption{Execution time in seconds}
\label{table:time}
\cite{wang2021decoupling}.
\end{table}
\newpage
First three implementations are following Wang et al. baseline design \cite{wang2021decoupling}.

\chapter{Conclusion}
\label{ch6}
The conclusion on findings of the study, as well as some suggestions for further research, are summarized in this last chapter.
\section{Discussion}
The study was started with several simple questions:
Is it possible to build a more powerful GNN encoder?
How combining different GNN encoders might behave?
For the first question we showed that combining the GIN and GAT encoder showed improvement, a different combination of GIN and GCN was experimented but did not result in better result.
The results of GIN and GAT combined for representation of graphs showed better representation than using any experimented pure GNN encoder  this might be due to that GIN GNN encoders  demonstrated a way to more accurately represent aggregation of neighbors and GAT makes it possible to (implicitly) provide varying weights to the individual nodes that make up a neighborhood.
Another question was asked and experimented would changing the clustering method for the deep cluster infomax lead to better results? for this study Affinity Propagation was tested but unfortunately it turned out to be  computationally more expensive than k-means, takes more time for same tasks and results in lower performance.

\section{Future Work}
For Future work I had three more ideas that need to be examined singly or simultaneously namely, Enhancing GNN encoders combinations, Changing the classifier, and changing the k-means clustering used for the pre-training.
\subsection{Enhancing GNN encoders}
The improvement of a GNN encoder might be achieved with better combinations of different encoders, for this study GIN and GAT has shown better performance when their embedding results are averaged, other aggregations i.e. instead of averaging having weighted averaging or using maximum functions might lead to better results.
\subsection{Improving the pre-training}
An improvement of the pre-training might be achieved using different clustering algorithm other than k-Means, and finding a better way to determine the k number of clusters for k-means algorithm, an experiment with the affinity propagation to eliminate the need of k clusters defining did not show improvement.
\subsection{Classifier Improvement}
The classifier used is simple linear classifier on top of it sigmoid non-linearity, would using different classifier show better results? this question was not examined during this study, and can be investigated in future work.
\bibliography{bib}
\bibliographystyle{apalike}

\end{document}